\documentclass[]{pasj01}
\begin{document}
\Received{2017/01/10}%{yyyy/mm/dd}
\Accepted{2017/02/15}%{yyyy/mm/dd}
%\Published{yyyy/mm/dd}

\title{X-ray study of extended emission around M86 observed with Suzaku}

%%%% begin:list of authors
%% Do NOT capitalize all letters in "textsc".
%\author{Ukyo \textsc{Hishi}\altaffilmark{1}%
%%\thanks{Present Address is xxxxxxxxxx}
%}
%\altaffiltext{1}{Division of Mathematics and Physics, Graduate school of
%Natural Science and Technology, Kanazawa University,
%Kakuma-machi, Kanazawa, Ishikawa 920-1192}
%\email{hishi@astro.s.kanazawa-u.ac.jp}
%
%\author{Ryuichi \textsc{Fujimoto}\altaffilmark{2}}
%\altaffiltext{2}{Faculty of Mathematics and Physics, Kanazawa
%University, Kakuma-machi, Kanazawa, Ishikawa 920-1192}
%\email{fujimoto@se.kanazawa-u.ac.jp}
%
%\author{Misato \textsc{Kotake}\altaffilmark{1}}
%
%\author{Hiromasa \textsc{Ito}\altaffilmark{1}}
%
%\author{Keigo \textsc{Tanaka}\altaffilmark{1}}
%
%\author{Yu \textsc{Kai}\altaffilmark{1}}
%
%%%% end:list of authors

\author{Ukyo \textsc{Hishi}, Ryuichi \textsc{Fujimoto}, 
Misato \textsc{Kotake}, Hiromasa \textsc{Ito},
Keigo \textsc{Tanaka}, Yu \textsc{Kai}, Yuya \textsc{Kinoshita}
}
\affil{Faculty of Mathematics and Physics, Kanazawa
University, Kakuma-machi, Kanazawa, Ishikawa 920-1192}
\email{hishi@astro.s.kanazawa-u.ac.jp, fujimoto@se.kanazawa-u.ac.jp}

%% `\KeyWords{}' always has to be placed before `\maketitle'.
 %Do NOT move this preamble from here!

\maketitle

\begin{abstract}
We analyzed the Suzaku data of M86 and its adjacent regions to study the
extended emission around it. The M86 core, the plume, and the tail
extending toward the northwest were clearly detected, as well as the
extended halo around them. From the position angle $\sim\timeform{45D}$
to $\sim\timeform{275D}$, the surface brightness distribution of the
core and the extended halo was represented relatively well with a single
$\beta$-model of $\beta\sim0.5$ up to \timeform{15'}--\timeform{20'}.
The X-ray spectra of the core was represented with a two-temperature
model of $kT\sim 0.9$~keV and $\sim0.6$~keV. The
temperatures of the core and the halo have a positive gradient in the
center, and reach the maximum of $kT\sim1.0$~keV at
 $r\sim\timeform{7'}$, indicating that the halo gas
is located in a larger scale potential structure than that of the
galaxy.  The temperatures of the plume and the tail were
$0.86\pm0.01$~keV and $1.00\pm0.01$~keV. We succeeded in determining the
abundances of $\alpha$-element separately, for the core, the
plume, the tail, and the halo for the first time. Abundance ratios with
respect to Fe were consistent with the solar ratios everywhere, except
for Ne. The abundance of Fe was $\sim0.7$ in the core and in the plume,
while that in the tail was $\sim1.0$, but the difference was not
significant considering the uncertainties of the ICM. The abundance of
the halo was almost the same up to $r\sim\timeform{10'}$, and then it
becomes significantly smaller (0.2--0.3) at $r\gtrsim\timeform{10'}$,
indicating the gas with low metal abundance still remains in the outer
halo. From the surface brightness distribution, we
estimated the gas mass ($\sim3\times10^{10}M_\odot$) and the dynamical
mass ($\sim3\times10^{12}M_\odot$) in $r<100$~kpc. The gas mass to
the dynamical mass ratio was $10^{-3}$--$10^{-2}$, suggesting a
significant fraction of the halo gas has been stripped.
\end{abstract}

%%%%%%%%%%%%%%%%%%%%%%%%%%%%%%%%%%%%%%%%%%%%%%%%%%%%%%%%%%%%%%%%%%%%%%%%%
\section{Introduction}

%\noindent IMPORTANT NOTICE\\
%1. ``\verb|\draft|'' creates single column and double spaces format.\\
%2. If you comment out ``\verb|\draft|'', the output will be double column
%   and single space.\\
%3. For cross-references, the use of ``\verb|\label|, \verb|\ref|, \verb|\cite|'' 
%   and the thebibliography environment is strongly recommended. \\
%4. Do NOT use ``\verb|\def|, \verb|\renewcommand|''.\\
%5. Do NOT redefine commands provided by PASJ01.cls.\\
%

M86 (NGC~4406) is a bright elliptical galaxy in the Virgo cluster,
located about \timeform{1.26D}, or about 350~kpc in projection, from the
Virgo cluster center M87.  Its redshift is $z=-0.000747\pm0.000017$
\citep{Cappellari11}, i.e., it is approaching us with the line-of-sight
velocity of $224\pm5$~km\,s$^{-1}$. On the other hand, the redshift of
M87 is $z=0.004283\pm0.000017$, and it is going away from us with the
line-of-sight velocity of $1284\pm5$~km\,s$^{-1}$. It is also reported
that M86 is only about 1~Mpc more distance than M87 \citep{Mei07}.
Therefore, M86 is likely moving in the Virgo cluster with a relative
line-of-sight velocity of about 1500~km\,s$^{-1}$ with respect to the
intracluster medium (ICM). This is much larger than the velocity
dispersion of galaxies in the Virgo cluster ($\sim700$~km\,s$^{-1}$)
\citep{Binggeli99}, and hence, the direction of motion is considered
close to the line-of-sight direction. Since the sound speed is
730~km\,s$^{-1}$ for the cluster ICM of $kT=2$~keV, M86 is moving with a
Mach number of $\gtrsim2$.  M86 is thought to be the dominant member of
one of the subgroups within the Virgo cluster (e.g.,
\cite{Virgo_ROSAT,Schindler99}). Therefore, M86 provides a good
opportunity to study the interaction between the interstellar medium
(ISM) and the ICM as well as the interaction between the subcluster and
the ICM.

Characteristic features of M86 were reported by various authors,
especially in the X-ray band, since it is sensitive to the hot ISM in
the elliptical galaxies. \citet{Forman79} discovered a plume of soft
X-ray emission, which is thought to be stripped from M86 by ram-pressure
with the Virgo ICM (see also \cite{White91}). Using ROSAT data,
\citet{M86_ROSAT} showed that the temperatures of the galaxy and the
plume are both $\sim0.8$~keV. Using Chandra data, \citet{M86_Chandra}
discovered a very long tail toward northwest, of 150~kpc in projection
and the true length of $\gtrsim380$~kpc. They also detected a
discontinuity of the X-ray surface brightness, which was interpreted as
the density jump due to the shock.  M86 was also observed by XMM-Newton
\citep{M86_XMM1,M86_XMM2}. \citet{M86_XMM2} examined the temperature and
abundance distributions in detail, and reported the existence of cool
($\sim0.6$~keV) gas trailing to the northwest of M86 and, also to the
east of M86 in the direction of NGC~4438.

In this paper, we report the results of our analysis of the Suzaku
archival data of M86 and its adjacent regions, to study the extended
emission around M86. We adopt a distance to the Virgo cluster of
16.5~Mpc, which gives a scale of 4.8~kpc per \timeform{1'}. All error
ranges are 90\% confidence intervals, and the $F$-test significance
level is 1\%, unless otherwise stated.

%\newpage

%%%%%%%%%%%%%%%%%%%%%%%%%%%%%%%%%%%%%%%%%%%%%%%%%%%%%%%%%%%%%%%%%%%%%%%%%
\section{Data reduction}

\begin{table}
  \tbl{Datasets used in the analysis.}{%
  \begin{tabular}{lllll}
      \hline
      No. & ObsID & Object & Obs Date & Exposure (ks) \\
      \hline
      1 & 803043010 & NGC 4406 (M86) & 2009-06-19 & 102 \\
      2 & 808045010 & NGC 4438 Tail & 2013-12-10 & 103 \\
      3 & 800017010 & NGC 4388 & 2005-12-24 & 124 \\
      \hline
    \end{tabular}}\label{tab:obslog}
\end{table}

We used three datasets of Suzaku version 2.5 products, archived in Data
ARchives and Transmission System (DARTS) at ISAS/JAXA.  M86 was observed
on 2009 June 19. Adjacent pointings aiming at NGC~4438 and NGC~4388 were
also used. They are summarized in table~\ref{tab:obslog}.

\texttt{HEASoft 6.15.1} was used for data processing, extraction and
analysis.  The data were reprocessed using \texttt{aepipeline v1.1.0},
with \texttt{CALDB} version 20150105 for the dataset \#1 (ObsID
803043010) and \#2 (ObsID 808045010), and with \texttt{CALDB} version
20140520 for \#3 (ObsID 800017010), following the standard screening
criteria described in the Suzaku Data Reduction
Guide\footnote{http://heasarc.gsfc.nasa.gov/docs/suzaku/analysis/abc/}
Version 5.0.  There were four independent XIS units (XIS0--3), but XIS2
was inoperative since 2006 November 9, and hence, XIS2 data were not
available for the dataset \#1 and \#2. Event files of $5\times5$ and
$3\times3$ editing modes were combined per sensor after the
reprocessing. The regions which were illuminated by the calibration
sources were discarded.  The net exposure time of each observation is
summarized in table~\ref{tab:obslog}. The average count rate of XIS1 in
the 0.5--5~keV band was 2.0, 1.0, and 0.93~counts\,s$^{-1}$,
respectively. It was checked that there was no statistically
significant variation in the light curves of the cleaned data.

Contribution of the particle background (Non-X-ray Background; NXB) of
each XIS unit was estimated using \texttt{xisnxbgen} tool and data taken
when the satellite saw the night side of the Earth stored in the
\texttt{CALDB}, by sorting them by the cut-off rigidity values, and
properly weighting them by the exposure time ratio, based on the results
by \citet{NXB}.  The detector redistribution matrix files (RMFs) were
generated with \texttt{xisrmfgen}, using the appropriate calibration
files at the time of the observation. On the other hand, responses of
the X-ray telescopes were implemented into ancillary response files
(ARFs), using ray-tracing based generator \texttt{xissimarfgen}
\citep{xissimarfgen}.  Time- and position-dependent contamination in the
optical path of each sensor was also considered by \texttt{xissimarfgen}.
When the ARFs were generated, we assumed a uniform source distribution
in a circle of \timeform{20'} radius.

%%%%%%%%%%%%%%%%%%%%%%%%%%%%%%%%%%%%%%%%%%%%%%%%%%%%%%%%%%%%%%%%%%%%%%%%%
\section{Image analysis and results}
\label{sec:image_analysis}

\begin{figure*}
 \begin{center}
  \includegraphics[width=8cm]{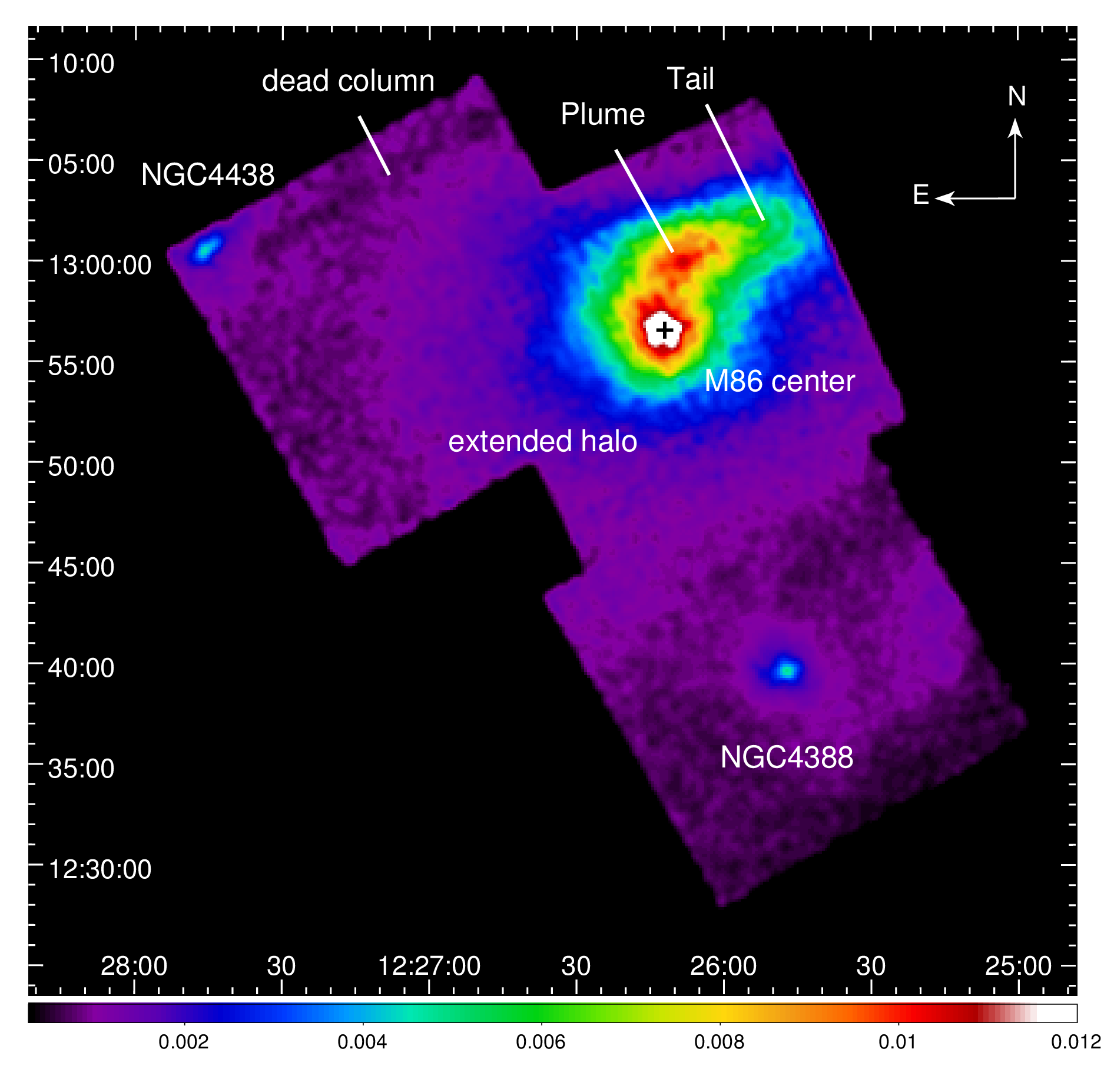}
  \includegraphics[width=8cm]{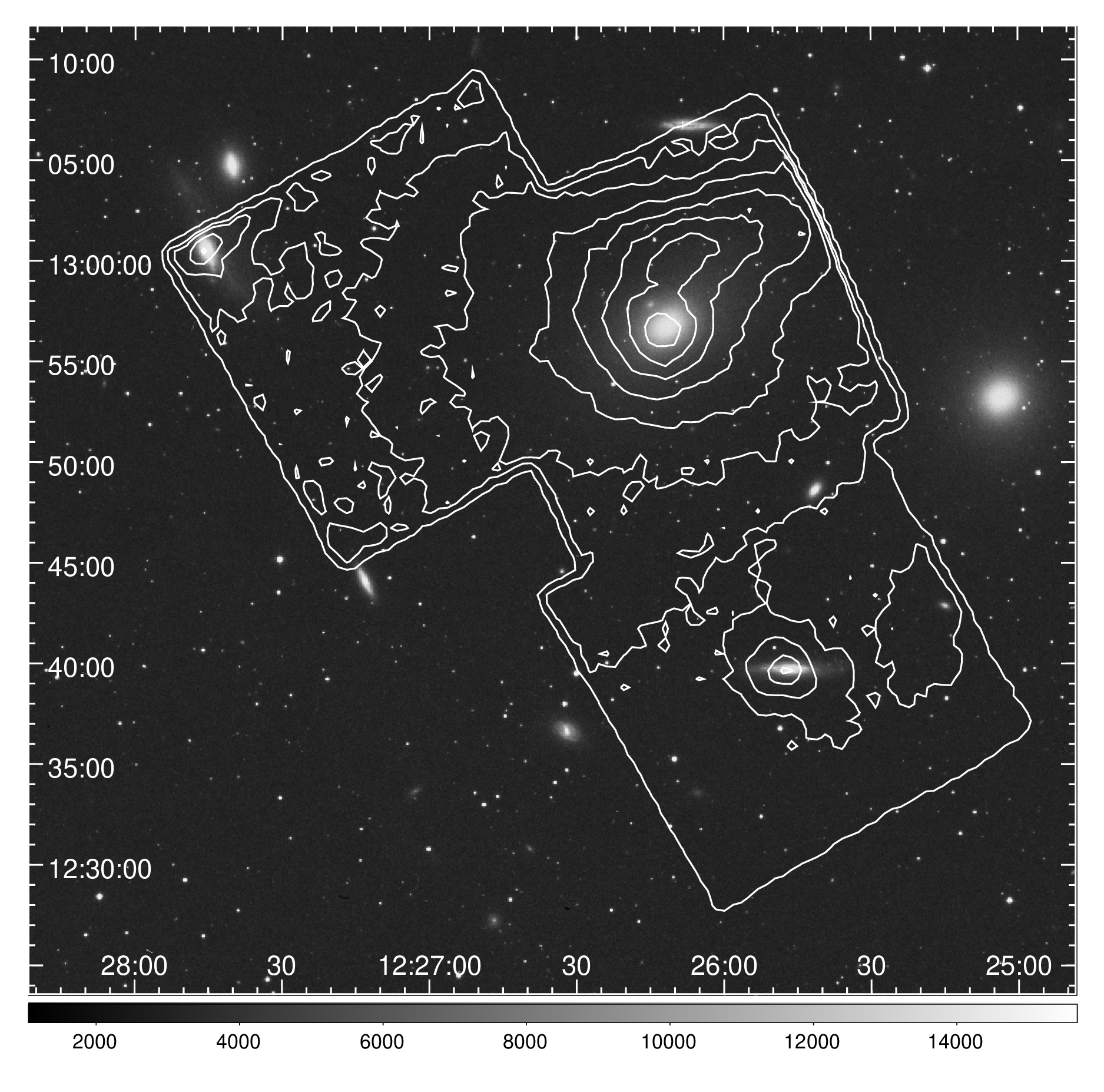}
 \end{center}
\caption{(Left) Background-subtracted, vignetting-corrected mosaic of
XIS images around M86 in the 0.8--1.2~keV energy band. Images of XIS0,
1, 3 were combined, and it was smoothed with a Gaussian of $\sigma$
corresponding to \timeform{0.42'}. (Right) X-ray contour map overlaid
with an optical image. The optical image was taken from the Digitized
Sky Survey.}
\label{fig:image}
\end{figure*}

A mosaic of the three pointings of the XIS in the 0.8--1.2~keV energy
band is shown in figure~\ref{fig:image}.  The energy range was selected
to be sensitive to the hot gas of $kT\sim1$~keV.  Images in this energy
band were extracted from the event files of XIS0, 1, 2, 3. They were
rebinned by a factor of 8 (\timeform{0.14'}), and were combined.  The
corresponding non-X-ray background images were generated using
\texttt{xisnxbgen}, and they were subtracted.  Then, flat field images
were generated, and the vignetting of the X-ray telescopes was corrected
by dividing the XIS images with the flat field images.  The mosaic was
generated, the corresponding exposure map was generated using
\texttt{xisexpmapgen}, and the mosaic was divided by the exposure map.
Finally, it was smoothed with a Gaussian of $\sigma=\timeform{0.42'}$ (3
rebinned pixels).

As clearly seen in figure~\ref{fig:image}, an extended emission of a
characteristic shape with two peaks was detected at the location of M86,
together with two more relatively weak sources at the position of
NGC~4388 and NGC~4438.  The brightest peak is the M86 center.  On the
north side of it, a large plume of emission is seen, and an elongated
tail extends toward the northwest.  An extended halo of the X-ray
emission is seen around the center of M86, which extends near NGC~4388
and NGC~4438.  All these are consistent with the previously reported
structures with high spatial resolution by \textit{ROSAT}
\citep{M86_ROSAT}, \textit{Chandra} \citep{M86_Chandra} and
\textit{XMM-Newton} \citep{M86_XMM2} observations. Note that
\citet{M86_Chandra} showed that the length of the tail in the plane of
the sky is 150~kpc (\timeform{0.51D}), but only a part of it was covered
with the XIS field of view.

\begin{figure*}
 \begin{center}
 \includegraphics[width=7.5cm]{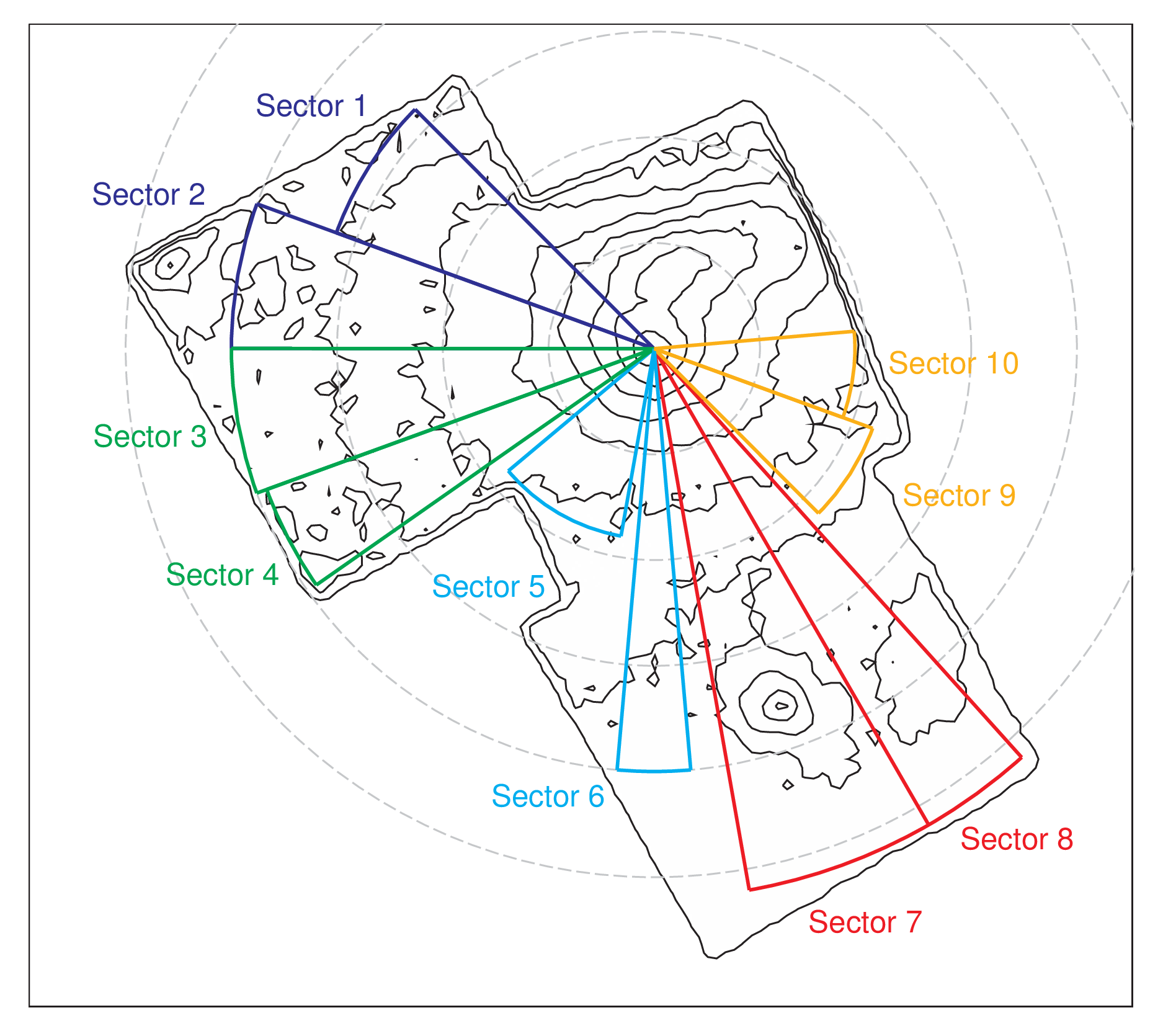}
 \includegraphics[width=8.5cm]{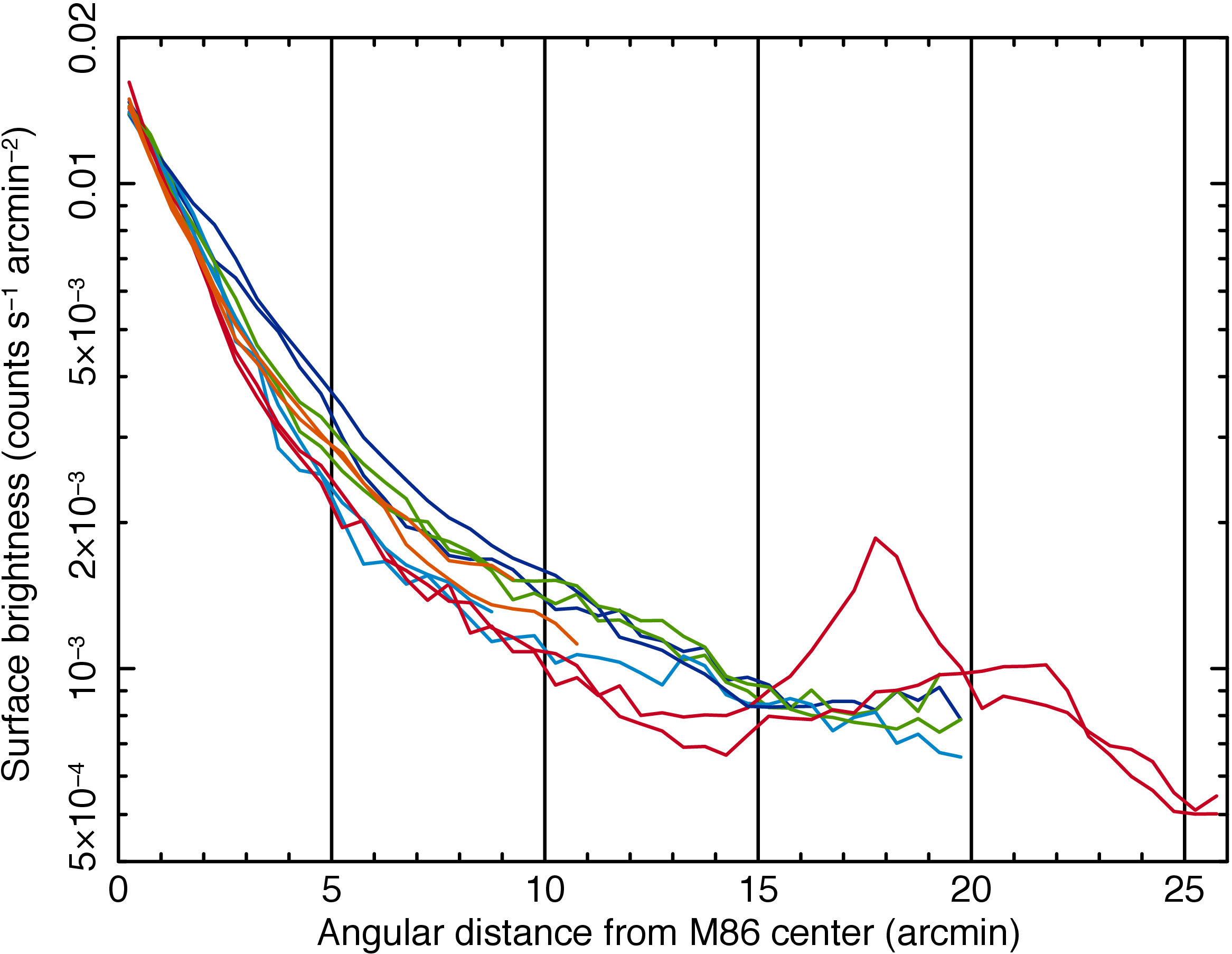}
 \end{center}
 \caption{(Left) Definition of sectors 1--10, to examine the surface
brightness profiles. Dotted circles show the distance from the M86
center, 5, 10, 15, 20, 25 arcmin, respectively. (Right) Surface
brightness profiles of sectors 1--10. Note that sector 7 contains
NGC~4388, and sector 8 contains an X-ray clump located near NGC~4388. }
\label{fig:sb}
\end{figure*}

Figure~\ref{fig:sb} shows definition of the ten sector regions centered
at the M86 center and surface brightness profiles of these regions.
Among these, sector 1 is the brightest, while sectors 6--8 are the
faintest. Sectors 7 and 8 contain NGC~4388 and an X-ray clump located
near NGC~4388, respectively, and hence, the surface brightness profiles
are complex. See section~\ref{sec:clump} for the X-ray clump.
Except for them, the profiles are similar.  In the eastern
regions (sectors 1--4), the slopes of the surface brightness change at
around \timeform{15'}, and they become flatter outside it.  In the the
southern regions (sectors 6--8), on the other hand, the slopes change at
\timeform{10'}--\timeform{12'}. Then, there is a contribution of
NGC~4388.  Outside $\sim\timeform{25'}$ (sectors 7 and 8), the surface
brightness becomes very low. At $r\gtrsim\timeform{15'}$ in the eastern
regions and $r\gtrsim{25'}$ in the southern regions, it appears that the
ICM of the Virgo cluster becomes dominant. Note that a factor of 2--3
difference in the ICM flux was reported around the M86 region
\citep{M86_ROSAT}. The surface brightness variation of these regions is
qualitatively consistent with their results.

\begin{table*}
\tbl{Best-fit parameters of the $\beta$ model fit for sectors 2, 3, 4,
and 6.}{
\begin{tabular}{cccccc}
\hline 
Parameter & Unit & Sector 2 & Sector 3 & Sector 4 & Sector 6 \\
\hline 
$\beta$   &        
& $0.51\pm0.03$ & $0.49\pm0.01$ & $0.44\pm0.01$ & $0.46\pm0.03$ \\
$\theta_0$& (arcmin) 
& $2.7\pm0.3$    & $2.7\pm0.2$ & $1.8\pm0.2$ & $1.6\pm0.2$ \\
$S_0$    & ($\times10^{-2}$ c\,s$^{-1}$\,arcmin$^{-2}$) 
& $1.20\pm0.06$ & $1.06\pm0.06$ & $1.28\pm0.10$ & $1.48_{-0.13}^{+0.14}$ \\
constant  & ($\times10^{-4}$ c\,s$^{-1}$\,arcmin$^{-2}$) 
& $6.3\pm0.5$ & 6.0 (fixed) & 6.0 (fixed) & $5.7\pm0.5$ \\
\hline
\end{tabular}}
\label{tab:sb_fit}
\end{table*}

\begin{figure*}
 \begin{center}
\includegraphics[width=80mm]{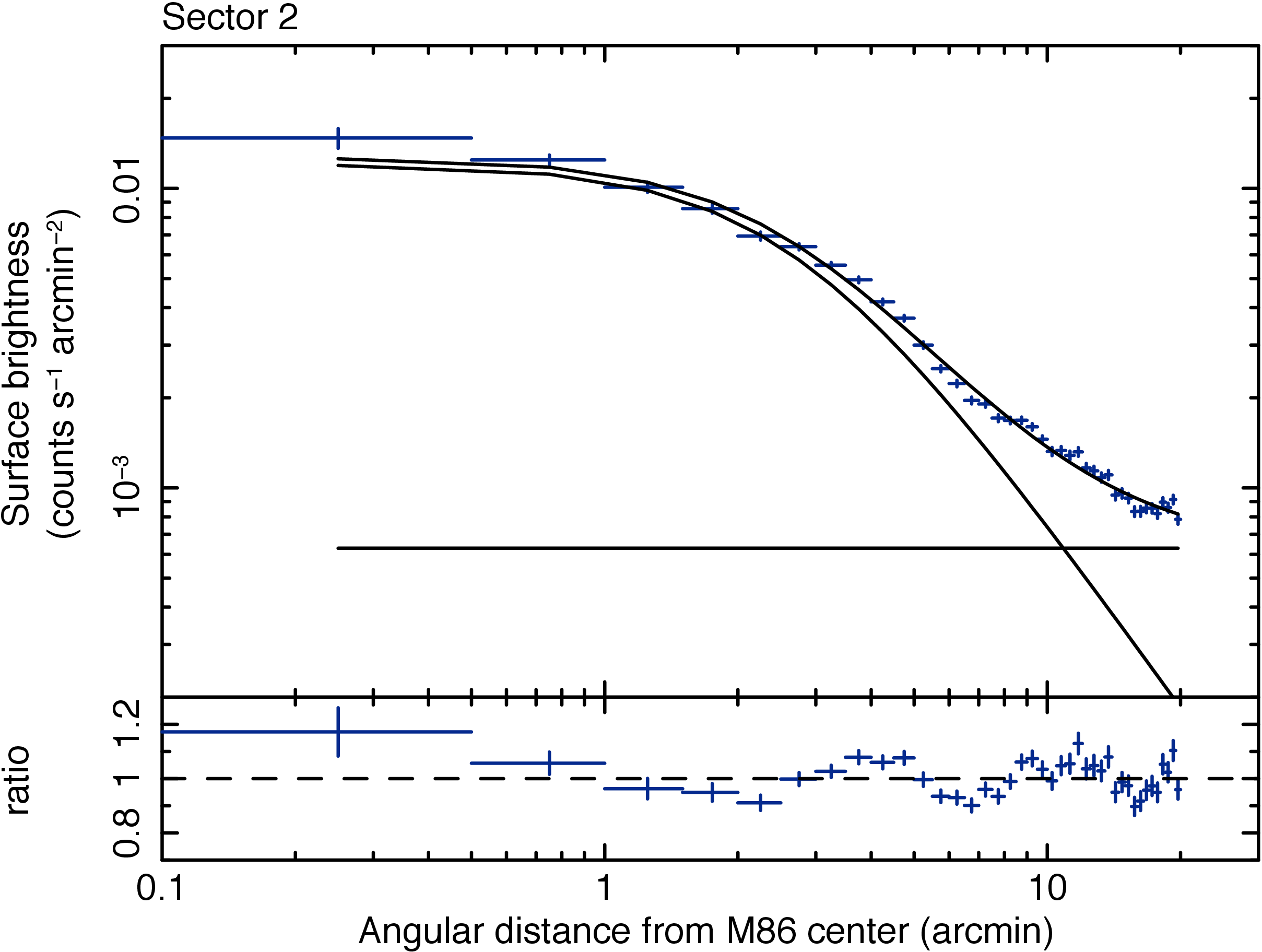}
\includegraphics[width=80mm]{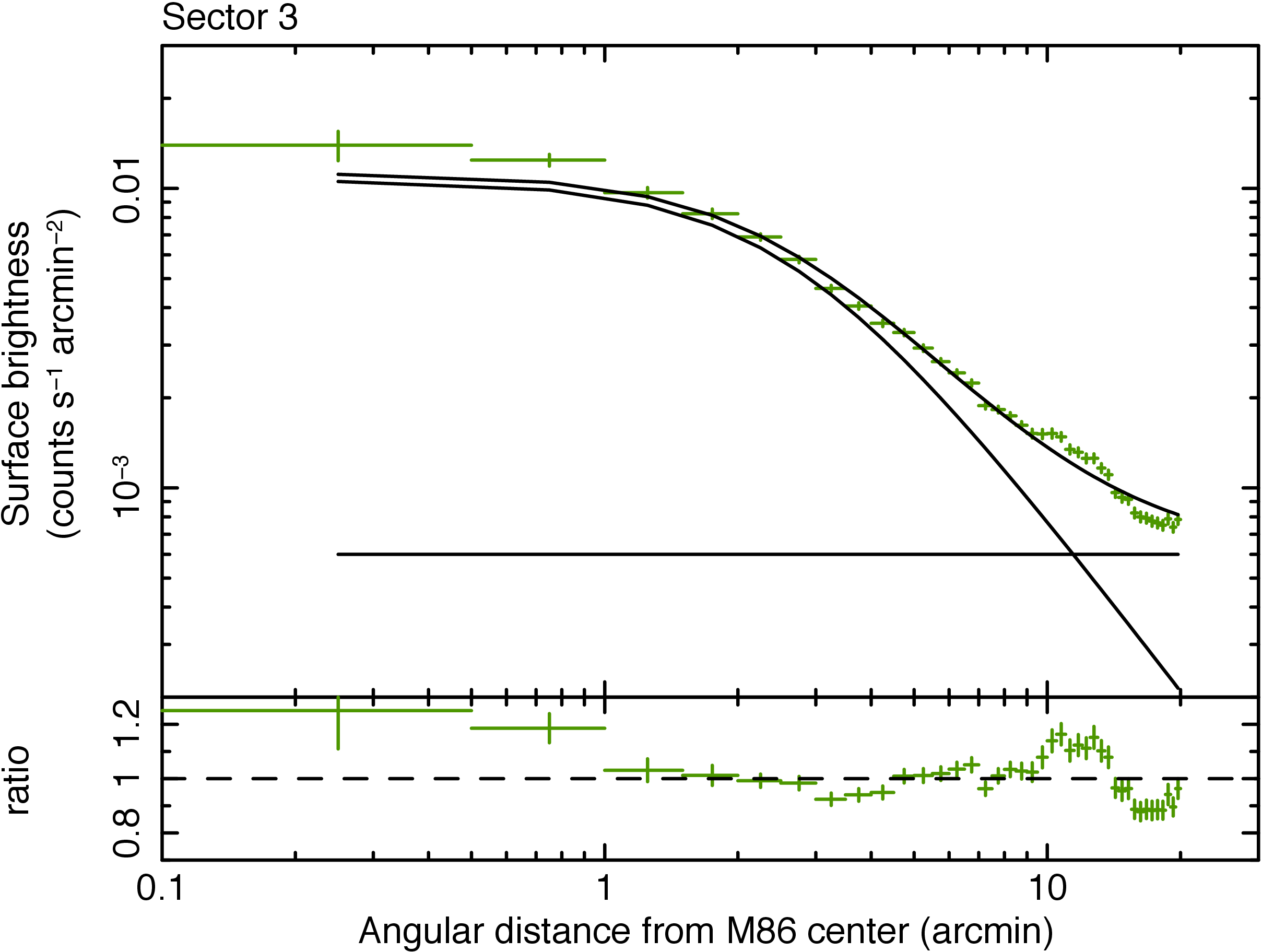}
\includegraphics[width=80mm]{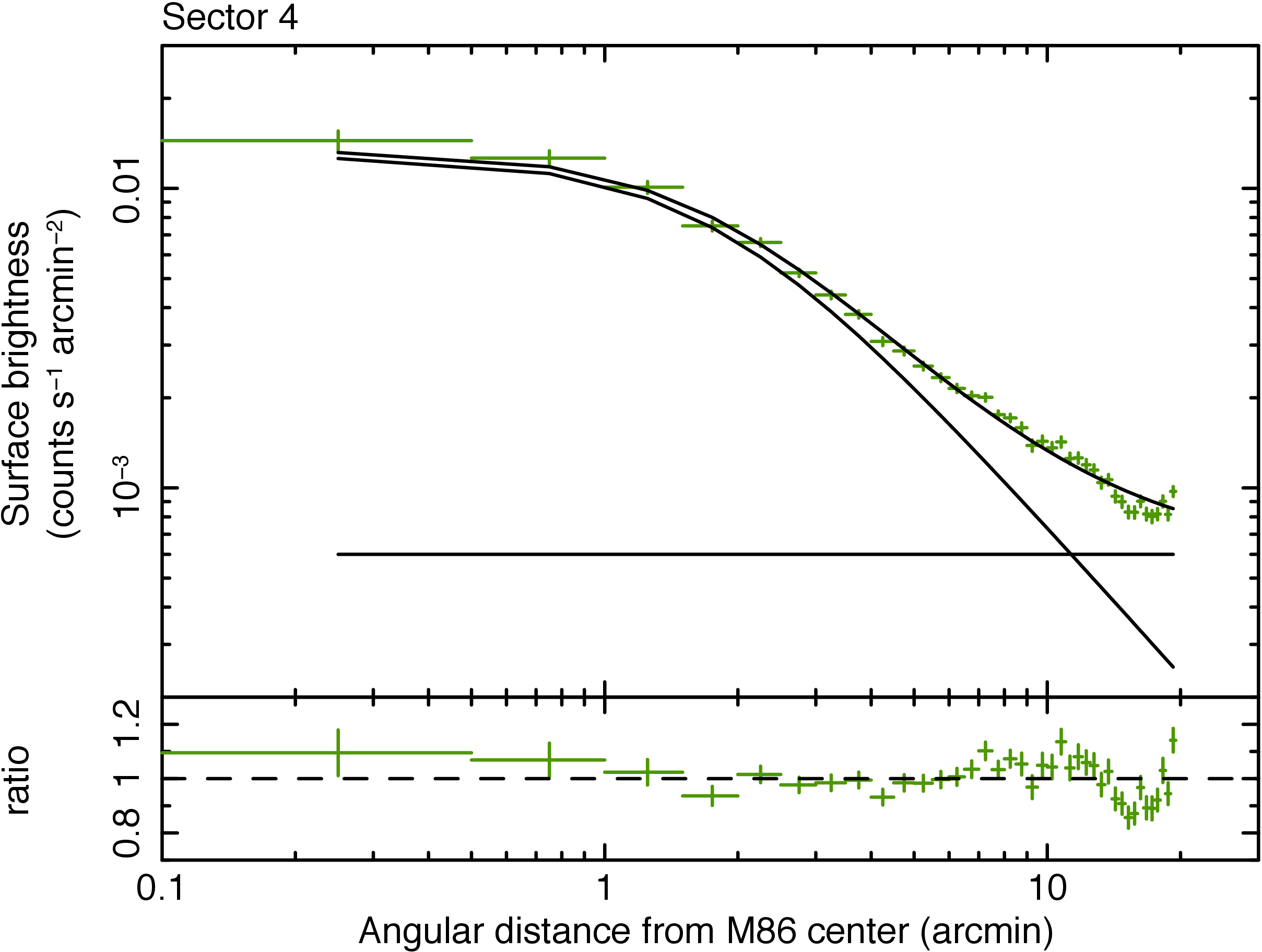}
\includegraphics[width=80mm]{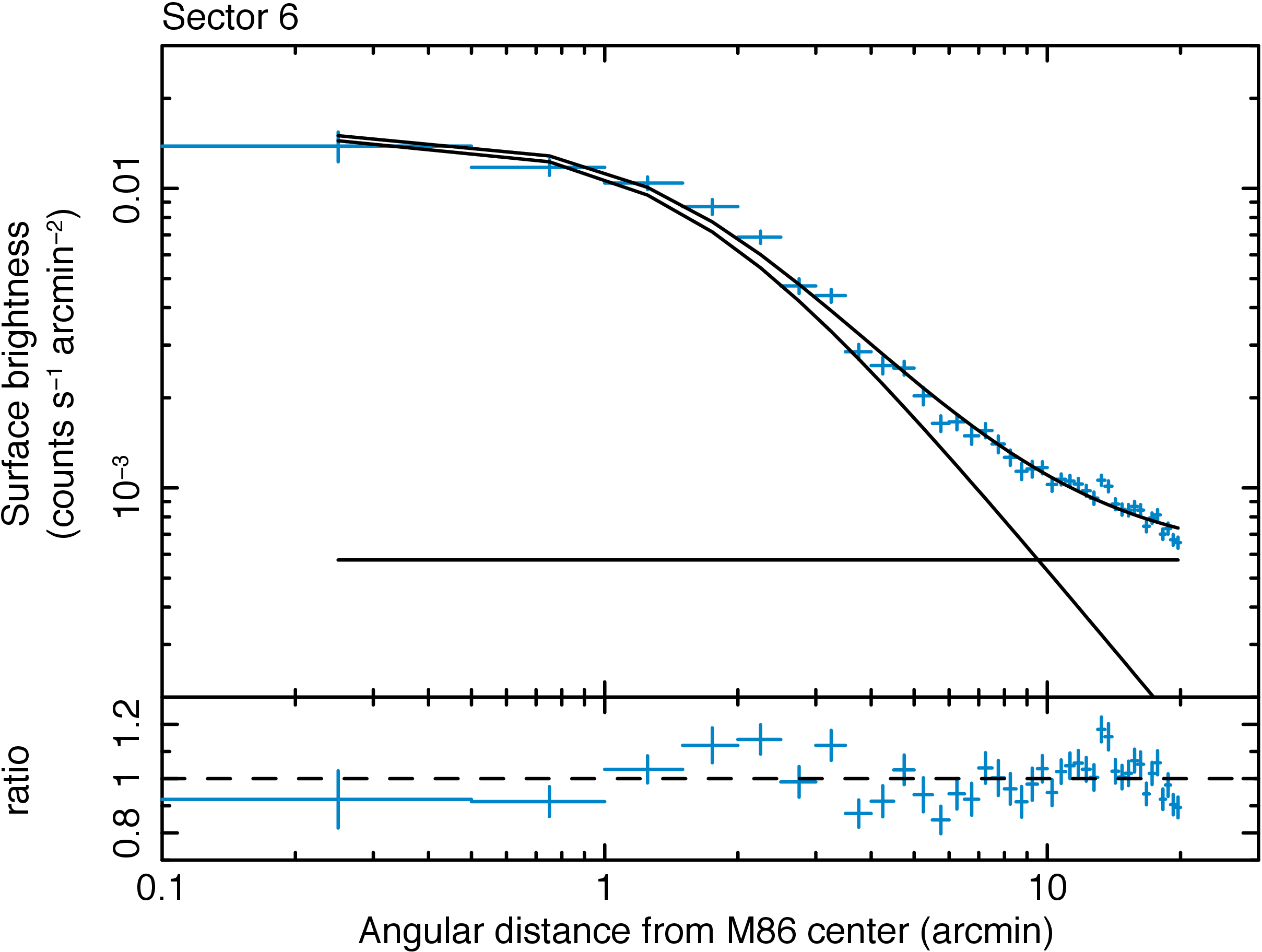}
 \end{center}
 \caption{The best fit models for the surface brightness profiles of
sectors 2, 3, 4, and 6.} \label{fig:sb_fit}
\end{figure*}

To understand the surface brightness profiles more quantitatively, we
fitted the profiles of sectors 2, 3, 4, and 6, with a $\beta$ model
\citep{beta_model}
\begin{equation}
 S(\theta) = S_0 \left\{
1+\left(\frac{\theta}{\theta_0}\right)^2
\right\}^{-3\beta+\frac{1}{2}}+{\rm constant},
\end{equation}
where a constant was introduced to represent the ICM and other
background and foreground components. The results are summarized in
table~\ref{tab:sb_fit} and the best fit models for sectors 2, 3, 4, and
6 are shown in
figure~\ref{fig:sb_fit}. Since it was not possible to constrain the
constant for sectors 3 and 4, it was fixed at the average value of
sectors 2 and 6. The surface brightness of the Suzaku image can be
represented relatively well with a single $\beta$-model of
$\beta\sim0.5$, up to \timeform{15'}--\timeform{20'} in the eastern and
southern regions.

%%%%%%%%%%%%%%%%%%%%%%%%%%%%%%%%%%%%%%%%%%%%%%%%%%%%%%%%%%%%%%%%%%%%%%%%%
\section{Spectral analysis and results}

\begin{figure}
 \begin{center}
\includegraphics[width=8cm]{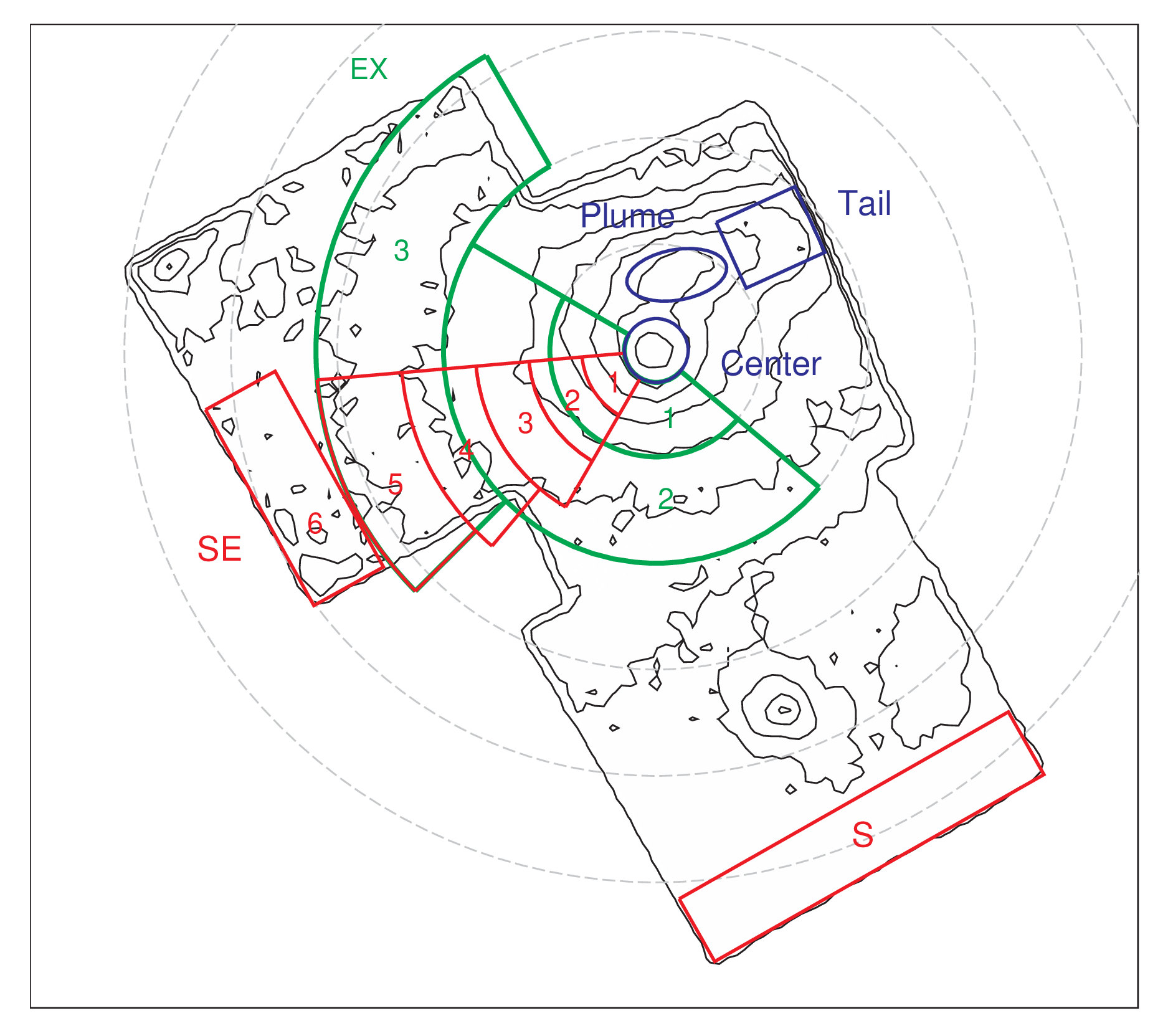}
%\vspace*{5cm}
 \end{center}
 \caption{Regions for spectral analysis. Dotted circles show the
distance from the M86 center, 5, 10, 15, 20, 25 arcmin, respectively.}
\label{fig:spectral_regions}
\end{figure}

As described in the previous section, the mosaic of the Suzaku XIS
images showed structures of the X-ray emission from M86, i.e., emissions
from the M86 core, the plume, the tail, and a diffuse emission around
them.  We defined regions as shown in figure~\ref{fig:spectral_regions},
and performed a model fitting to the spectral data extracted from these
regions. As an emission model from an optically-thin thermal plasma in
collisional ionization equilibrium, we used the APEC (Astrophysical
Plasma Emission Code) model \citep{APEC}, v2.0.2. It was used for the
galaxy hot gas, the ICM, and the galactic foregrounds, i.e., the Local
Hot Bubble (LHB) and the Milky Way Halo (MWH).  The temperatures of the
LHB and MWH were fixed at 0.11~keV (LHB) and 0.3~keV (MWH),
respectively.  The solar abundances ($Z$) by \citet{Lodders03} were
adopted in the fitting, and those of LHB and MWH were assumed the same
as the solar values ($1Z_\odot$).  The Cosmic X-ray Background (CXB) was
modeled with a power law function of a photon index 1.4 and a
normalization corresponding to 9.7 photons\,s$^{-1}$\,cm$^{-2}$ at 1~keV
\citep{CXB}. The galactic hydrogen column density was fixed at
$2.84\times10^{20}$~cm$^{-2}$ \citep{NH}.  Note that a constant ratio
between the BI CCD data and the FI CCD data were introduced. In all the
fitting, it was in the acceptable range ($\sim1.0\pm0.2$).

%%%%%%%%%%%%%%%%%%%%%%%%%%%%%%%%%%%%%%%%%%%%%%%%%%%%%%%%%%%%%%%%%%%%%%%%%
\subsection{Outer regions}
\label{sec:Outer_regions}

\begin{table*}
\tbl{Best fit parameters for the spectra of the outer regions SE6 and S.}{
\begin{tabular}{ccccccc}
\hline 
& & & \multicolumn{2}{c}{Region SE6} & \multicolumn{2}{c}{Region S}\\
Component & Parameter & Unit & $1T$ model & $2T$ model & $1T$ model & $2T$ model\\
\hline
APEC1 & $kT_{\rm}$ & (keV)
 & 1.887$_{-0.101}^{+0.098}$ & 2.089$_{-0.157}^{+0.231}$
 & 1.476$_{-0.047}^{+0.050}$ & 1.711$_{-0.129}^{+0.128}$
\\
& $Z$ & (solar)
 & 0.199$_{-0.051}^{+0.063}$ & 0.273$_{-0.079}^{+0.107}$
 & 0.182$_{-0.025}^{+0.033}$ & 0.265$_{-0.051}^{+0.055}$
\\
& Norm & ($\times 10^{-2}$)
 & 4.191$_{-0.272}^{+0.287}$ & 3.586$_{-0.384}^{+0.400}$
 & 2.436$_{-0.131}^{+0.113}$ & 1.784$_{-0.173}^{+0.323}$
\\
APEC2 & $kT$ & (keV)
 & -- & 0.922$_{-0.230}^{+0.346}$
 & -- & 1.018$_{-0.068}^{+0.060}$
\\
& $Z$ & (solar)
 & \multicolumn{4}{c}{$={\rm APEC1:}Z$}
\\
& Norm & ($\times 10^{-3}$)
 & -- & 3.327$_{-1.598}^{+5.753}$
 & -- & 3.984$_{-0.799}^{+1.789}$
\\
APEC (LHB) & $kT$ & (keV) & \multicolumn{4}{c}{0.11 (fixed)}\\
& Norm & ($\times 10^{-3}$)
 & 8.138$_{-2.352}^{+2.382}$ & 10.332$_{-2.616}^{+2.638}$
 & 1.574$_{-0.750}^{+0.754}$ & 2.543$_{-0.746}^{+0.824}$
\\
APEC (MWH) & $kT$ & (keV) & \multicolumn{4}{c}{0.3 (fixed)}\\
& Norm & ($\times 10^{-4}$) 
 & 8.567$_{-5.056}^{+5.036}$ & 4.083$_{-4.083}^{+6.685}$
 & 0.427$_{-0.427}^{+2.043}$ & $<1.534$
\\
-- & $\chi^2/\textrm{d.o.f.}$ & 
 & 333.37/279 & 320.99/277
 & 494.53/424 & 472.16/422
\\
%\multicolumn{2}{c}{F-value/probability}
% & -- & 5.342/0.00529
% & -- & 3.834/0.051
%\\
\hline
\end{tabular}}
\label{tab:outer_region_params}
\begin{tabnote}
The abundances of LHB and MWH were fixed at 1 solar.  The normalizations
of the APEC components are in units of
$\frac{10^{-14}}{4\pi[D_A(1+z)]^2}\int n_en_H dV$ per $400\pi$
arcmin$^2$, where $D_A$ is the angular diameter distance to the source
(cm), $n_e$ and $n_H$ are the electron and hydrogen densities
(cm$^{-3}$).
\end{tabnote}
\end{table*}

\begin{figure*}
\begin{center}
\includegraphics[width=80mm]{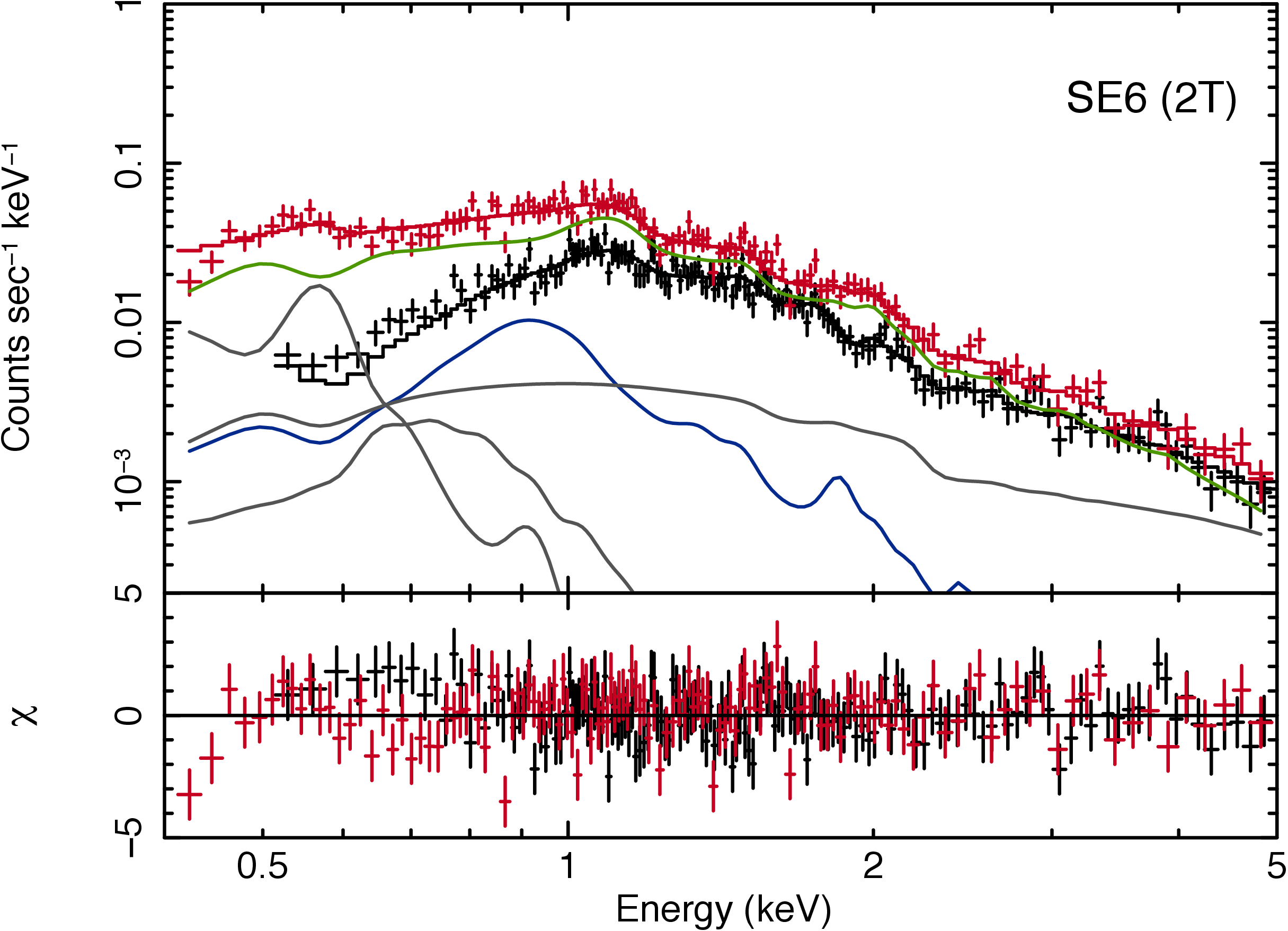}
\includegraphics[width=80mm]{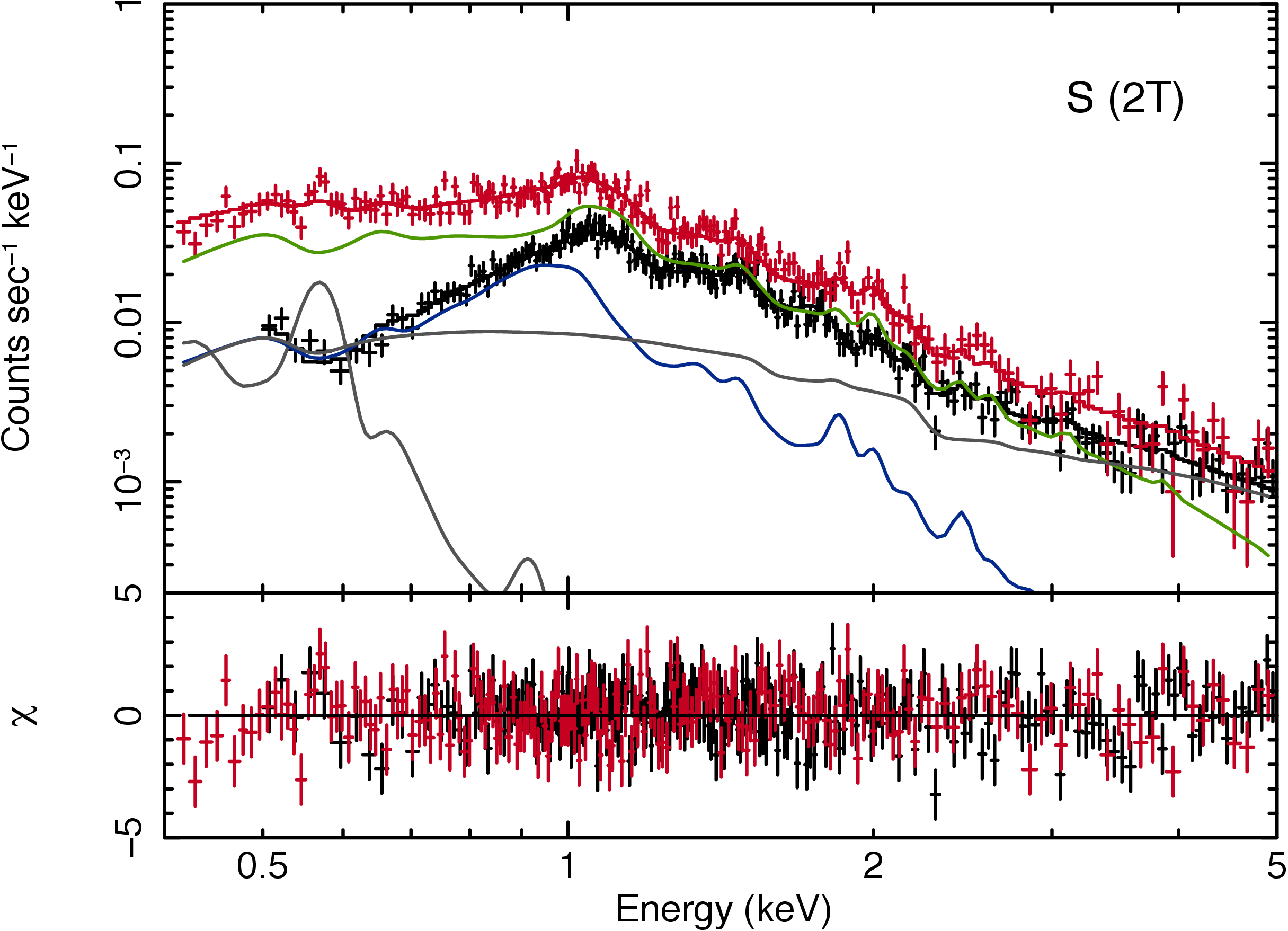}
\end{center}
\caption{Spectra of the outer regions SE6 and S, and the best-fit
models.  The red and black crosses show the data points of BI and FI CCD
data, and the red and black solid curves are the best fit models for
them.  The green, magenta, and gray curves are the high-$T$ component,
the low-$T$ component, and the backgrounds/foreground components (CXB,
LHB, MWH), respectively.  Only the components for the BI model are shown.
} \label{fig:outer_region_spectra}
\end{figure*}

First, we analyzed spectra of outer regions to the south-east and the
south of M86 center, SE6 and S, respectively, to evaluate the ICM around
M86.  One or two APEC models were employed to represent the emission in
these regions, in addition to two APEC models for the Galactic
components (LHB and MWH) and a power law model for the CXB. When two
APEC models were employed, their abundances were linked. The spectra and
the best fit models are shown in figure~\ref{fig:outer_region_spectra},
and the best fit parameters are summarized in
table~\ref{tab:outer_region_params}.  Fitting was significantly improved
by employing two APEC models.  The $F$-test provided the probability of
$5.3\times10^{-3}$ for region SE6 and $5.7\times10^{-5}$ for region S,
respectively. In both cases, the higher temperature component was
dominant. The temperature was $kT=2.09^{+0.23}_{-0.16}$~keV and
$1.71\pm0.13$~keV, respectively, and the abundance was
$\sim0.27Z_\odot$. The temperature of the other component was
$kT\sim1$~keV. Since the higher temperature component was dominant, and
its temperature was $\sim2$~keV, it was interpreted as the ICM
emission. The lower temperature component was, on the other hand,
considered a contribution of the extended emission of M86.
Note that the normalizations of the LHB and the MWH were significantly
different between the two regions, and their error bars were large. We
compared them with those shown in \citet{Simionescu15}, who
determined the spectral parameters of the LHB and the MWH 
using a set of 12 ROSAT All-sky Survey data beyond the virial radius of the Virgo cluster.
After the unit
conversion, the normalization of the MWH of region S was consistent with
\citet{Simionescu15} within an error range, while that of region SE6 was
larger by a factor of $\sim6$. The LHB normalizations were larger by
about an order even for region S. We also compared the normalizations
with those reported by \citet{Yoshino09}, who studied soft X-ray
diffuse foreground emission with Suzaku. The normalizations of our
results were within the variation range of those shown in \citet{Yoshino09},
except for the LHB normalization of region SE6. The
foreground components of region SE6 might have been affected by solar
activities.

A region around SE6 was observed with \textit{ROSAT} and
\textit{Chandra}.  According to \citet{M86_ROSAT}, the ICM temperature
of the \textit{ROSAT} South East quadrant was 1.76~keV, while
\citet{M86_Chandra} reported that the spectrum of their R18 region was
represented with $kT=1.085$~keV and 2.107~keV APEC models. These were
close to the temperatures of the $1T$ model and the $2T$ model shown in
table~\ref{tab:outer_region_params}, respectively. Therefore, we judged
that the results were in agreement with each other.  On the other hand,
the ICM temperature of \textit{ROSAT} South West quadrant was 2.09~keV
\citep{M86_ROSAT}, which was not consistent with the temperatures of our
region S. In the following analysis, the $2T$ model parameters for
region SE6 was regarded as representative of the ICM emission, and the
abundance of the ICM was assumed to be $0.27Z_\odot$. It is consistent
with the number (26\%) adopted by \citet{M86_Chandra}.  Note that the
metallicity of the Virgo cluster at the same radius from M87 was also
reported to be $Z\sim0.3Z_\odot$ \citep{VC_XMM,M86_XMM2}.

%%%%%%%%%%%%%%%%%%%%%%%%%%%%%%%%%%%%%%%%%%%%%%%%%%%%%%%%%%%%%%%%%%%%%%%%%
\subsection{M86 center and halo regions}

Secondly, the M86 center and the halo regions were analyzed. The regions
used for the analysis were the center and 5 annular sectors (SE1--5,
EX1--3) shown in figure~\ref{fig:spectral_regions}.  The center region
is a circle of \timeform{1.5'} radius at the M86 center. This is smaller
than $\theta_0$ shown in table~\ref{tab:sb_fit}, and surface brightness
can be regarded approximately constant in this radius.  Annular sectors
SE1--5 were defined from the position angle \timeform{95D} to
\timeform{150D}, to avoid contamination from NGC~4438 and NGC~4388. The
radius of the five regions were \timeform{1.5'}--\timeform{3.5'},
\timeform{3.5'}--\timeform{6'}, \timeform{6.0'}--\timeform{8.5'},
\timeform{8.5'}--\timeform{12'}, and \timeform{12'}--\timeform{16'},
respectively.  The dataset \#1 was used for the inner three regions
(SE1--3), while the dataset \#2 was used for the outer two regions (SE4,
5). To examine the abundance distribution, wider regions were needed
from the statistical point of view. Hence, we also defined annular
sectors EX1--3 for this purpose. The radius of the three regions were
\timeform{1.5'}--\timeform{5'}, \timeform{5'}--\timeform{10'}, and
\timeform{10'}--\timeform{16'}. The dataset \#1 was used for EX1 and
EX2, while the dataset \#2 was used for EX3.

%%%%%%%%%%%%%%%%%%%%%%%%%%%%%%%%%%%%%%%%%%%%%%%%%%%%%%%%%%%%%%%%%%%%%%%%%
\begin{table*}
\tbl{Best-fit spectral parameters for the center and SE regions obtained
from the $1T$ fit and the $2T$ fit.}{
\begin{tabular}{ccccccccc}
\hline 
Component & Parameter & Unit & Center &SE1 & SE2 & SE3 & SE4 & SE5 \\
\hline
\multicolumn{9}{c}{$1T$ model}\\
\hline
vAPEC & $kT$&(keV) & 0.800$_{-0.009}^{+0.009}$ &0.813$^{+0.016}_{-0.016}$ &0.933$^{+0.022}_{-0.026}$ &0.984$^{+0.022}_{-0.026}$ &0.957$^{+0.033}_{-0.042}$ &0.872$^{+0.076}_{-0.089}$ \\
& $Z_{\rm O}$  &(solar) & 0.666$_{-0.213}^{+0.271}$ & 0.732$^{+0.486}_{-0.352}$ &0.919$^{+0.691}_{-0.463}$ &0.537$^{+1.014}_{-0.537}$ &0.236$^{+0.708}_{-0.236}$ &$<$0.846\\
& $Z_{\rm Ne}$ &(solar) & 2.683$_{-0.502}^{+0.666}$ &2.319$^{+0.976}_{-0.673}$ &1.480$^{+1.088}_{-0.797}$ &2.751$^{+2.601}_{-1.374}$ &0.964$^{+1.343}_{-0.964}$ &1.454$^{+2.518}_{-1.109}$ \\
& $Z_{\rm Mg}$ &(solar) & 0.810$_{-0.139}^{+0.183}$ &0.662$^{+0.275}_{-0.192}$ &0.696$^{+0.356}_{-0.234}$ &0.976$^{+0.806}_{-0.398}$ &0.387$^{+0.370}_{-0.244}$ &$<$0.246 \\
& $Z_{\rm Si}$ &(solar) & 0.646$_{-0.100}^{+0.128}$ &0.540$^{+0.191}_{-0.141}$ &0.552$^{+0.230}_{-0.161}$ &0.733$^{+0.502}_{-0.264}$ &0.266$^{+0.218}_{-0.178}$ &0.238$^{+0.320}_{-0.238}$ \\
& $Z_{\rm S}$  &(solar) & 0.896$_{-0.196}^{+0.228}$ & 0.526$^{+0.314}_{-0.280}$ &0.853$^{+0.410}_{-0.326}$ &0.695$^{+0.637}_{-0.486}$ &0.614$^{+0.496}_{-0.421}$ &0.207$^{+0.796}_{-0.207}$ \\
& $Z_{\rm Fe}$ &(solar) & 0.663$_{-0.097}^{+0.133}$ &0.561$^{+0.168}_{-0.110}$ &0.571$^{+0.229}_{-0.138}$ &0.707$^{+0.523}_{-0.229}$ &0.425$^{+0.261}_{-0.135}$ &0.248$^{+0.283}_{-0.099}$ \\
& Norm &($\times 10^{-2}$) & 16.288$_{-2.562}^{+2.573}$ &9.293$^{+1.992}_{-1.981}$ &4.253$^{+1.097}_{-1.093}$ &2.123$^{+0.851}_{-0.847}$ &1.883$^{+0.631}_{-0.628}$ &1.070$^{+0.425}_{-0.492}$ \\
%\hline
APEC (ICM) & $kT$ & (keV) & \multicolumn{6}{c}{2.1 (fixed)}\\
 & $Z$ & (solar) & \multicolumn{6}{c}{0.27 (fixed)}\\
 & Norm &($\times 10^{-2}$) & 2.720$_{-1.634}^{+1.628}$ &3.430$^{+0.625}_{-0.629}$ &2.962$^{+0.469}_{-0.475}$ &3.066$^{+0.424}_{-0.427}$ &2.934$^{+0.299}_{-0.302}$ &3.205$^{+0.238}_{-0.232}$ \\
%\hline
Power law (LMXB) & $\Gamma$& & 1.5 (fixed) & -- & -- & -- & -- & -- \\
& Norm&($\times 10^{-3}$) & 4.506$_{-1.355}^{+1.356}$ & -- & -- & -- & -- & -- \\
-- & $\chi^2/{\rm d.o.f.}$ && 671.48/640 &394.74/351&408.06/401&393.65/393&354.09/329&405.31/371\\
\hline
\multicolumn{9}{c}{$2T$ model}\\
\hline
vAPEC1 & $kT$&(keV) & 0.876$_{-0.036}^{+0.029}$ & 0.810$_{-0.016}^{+0.015}$ & 0.833$_{-0.228}^{+0.059}$ & 0.934$_{-0.084}^{+0.038}$ & 0.967$_{-0.035}^{+0.030}$ & 0.843$_{-0.843}^{+0.084}$ \\
& $Z_{\rm O}$  &(solar) & 0.647$_{-0.207}^{+0.266}$ & 0.841$_{-0.325}^{+0.457}$ & 0.991$_{-0.515}^{+0.808}$ & 0.698$_{-0.586}^{+1.081}$ & 0.106$_{-0.106}^{+0.351}$ & 0.122$_{-0.122}^{+0.849}$ \\
& $Z_{\rm Ne}$ &(solar) & 2.277$_{-0.526}^{+0.695}$ & 2.118$_{-0.536}^{+0.776}$ & 1.786$_{-0.998}^{+1.413}$ & 3.489$_{-1.603}^{+3.233}$ & 0.356$_{-0.356}^{+0.705}$ & 1.538$_{-0.562}^{+2.199}$ \\
& $Z_{\rm Mg}$ &(solar) & 0.874$_{-0.158}^{+0.213}$ & 0.700$_{-0.166}^{+0.125}$ & 0.856$_{-0.289}^{+0.466}$ & 1.165$_{-0.451}^{+0.981}$ & 0.290$_{-0.139}^{+0.182}$ & 0.087$_{-0.087}^{+0.385}$ \\
& $Z_{\rm Si}$ &(solar) & 0.715$_{-0.115}^{+0.150}$ & 0.581$_{-0.123}^{+0.172}$ & 0.673$_{-0.195}^{+0.300}$ & 0.807$_{-0.279}^{+0.585}$ & 0.243$_{-0.102}^{+0.110}$ & 0.332$_{-0.279}^{+0.315}$ \\
& $Z_{\rm S}$  &(solar) & 0.921$_{-0.205}^{+0.248}$ & 0.538$_{-0.216}^{+0.236}$ & 0.937$_{-0.342}^{+0.456}$ & 0.796$_{-0.403}^{+0.633}$ & 0.463$_{-0.234}^{+0.269}$ & 0.427$_{-0.432}^{+0.518}$ \\
& $Z_{\rm Fe}$ &(solar) & 0.745$_{-0.118}^{+0.162}$ & 0.635$_{-0.093}^{+0.169}$ & 0.700$_{-0.183}^{+0.320}$ & 0.832$_{-0.290}^{+0.657}$ & 0.234$_{-0.066}^{+0.117}$ & 0.339$_{-0.156}^{+0.327}$ \\
& Norm &($\times 10^{-2}$) & 12.097$_{-2.149}^{+2.078}$ & 8.227$_{-1.626}^{+1.682}$ & 2.466$_{-0.861}^{+0.959}$ & 1.554$_{-0.862}^{+0.863}$ & 3.339$_{-0.975}^{+1.001}$ & 0.801$_{-0.801}^{+0.479}$ \\
%\hline
vAPEC2 & $kT$&(keV) & 0.562$_{-0.059}^{+0.087}$ & 1.819$_{-0.324}^{+0.215}$ & 1.252$_{-0.647}^{+0.381}$ & 1.719$_{-0.481}^{+0.407}$ & $>5.582$ & 1.875 \\
& Norm &($\times 10^{-2}$) & 3.625$_{-1.264}^{+2.102}$ & 3.273$_{-1.450}^{+2.396}$ & 1.656$_{-0.608}^{+0.791}$ & 1.115$_{-0.596}^{+1.511}$ & 0.522$_{-0.179}^{+0.325}$ & 0.880$_{-0.880}^{+1.886}$ \\
APEC (ICM) & $kT$ & (keV) & \multicolumn{6}{c}{2.1 (fixed)}\\
 & $Z$ & (solar) & \multicolumn{6}{c}{0.27 (fixed)}\\
 & Norm &($\times 10^{-2}$) & 3.048$_{-2.085}^{+1.902}$ & 0 & 2.558$_{-1.279}^{+0.648}$ & 1.984$_{-1.984}^{+1.077}$ & 1.593$_{-0.831}^{+0.433}$ & 2.384$_{-2.198}^{+0.975}$ \\
%\hline
Power law (LMXB) & $\Gamma$& & 1.5 (fixed) & -- & -- & -- & -- & -- \\
& Norm&($\times 10^{-3}$) & 4.280$_{-1.485}^{+1.564}$ & -- & -- & -- & -- & -- \\
-- & $\chi^2/{\rm d.o.f.}$ && 656.77/638 & 379.20/349 & 395.73/399 & 380.85/391 & 343.36/327 & 403.16/369 \\
\hline
\end{tabular}}
\label{tab:SE_1T}
\begin{tabnote}
In all the cases, APEC models for LHB ($kT=0.11$~keV, $Z=1Z_\odot$,
$\textrm{Norm}=9.6\times10^{-3}$) and MWH ($kT=0.3$~keV, $Z=1Z_\odot$,
$\textrm{Norm}=1.9\times10^{-3}$), and a power law model for CXB
($\Gamma=1.4$, $\textrm{Norm}=1.063\times10^{-3}$) were included. The
normalizations of the APEC components are in units of
$\frac{10^{-14}}{4\pi[D_A(1+z)]^2}\int n_en_H dV$ per $400\pi$
arcmin$^2$, where $D_A$ is the angular diameter distance to the source
(cm), $n_e$ and $n_H$ are the electron and hydrogen number densities
(cm$^{-3}$). The normalizations of the power law are in units of
photons\,keV$^{-1}$\,cm$^{-2}$\,s$^{-2}$ at 1~keV per
$400\pi$~arcmin$^2$.
\end{tabnote}
\end{table*}

\begin{figure*}
\begin{center}
\includegraphics[width=80mm]{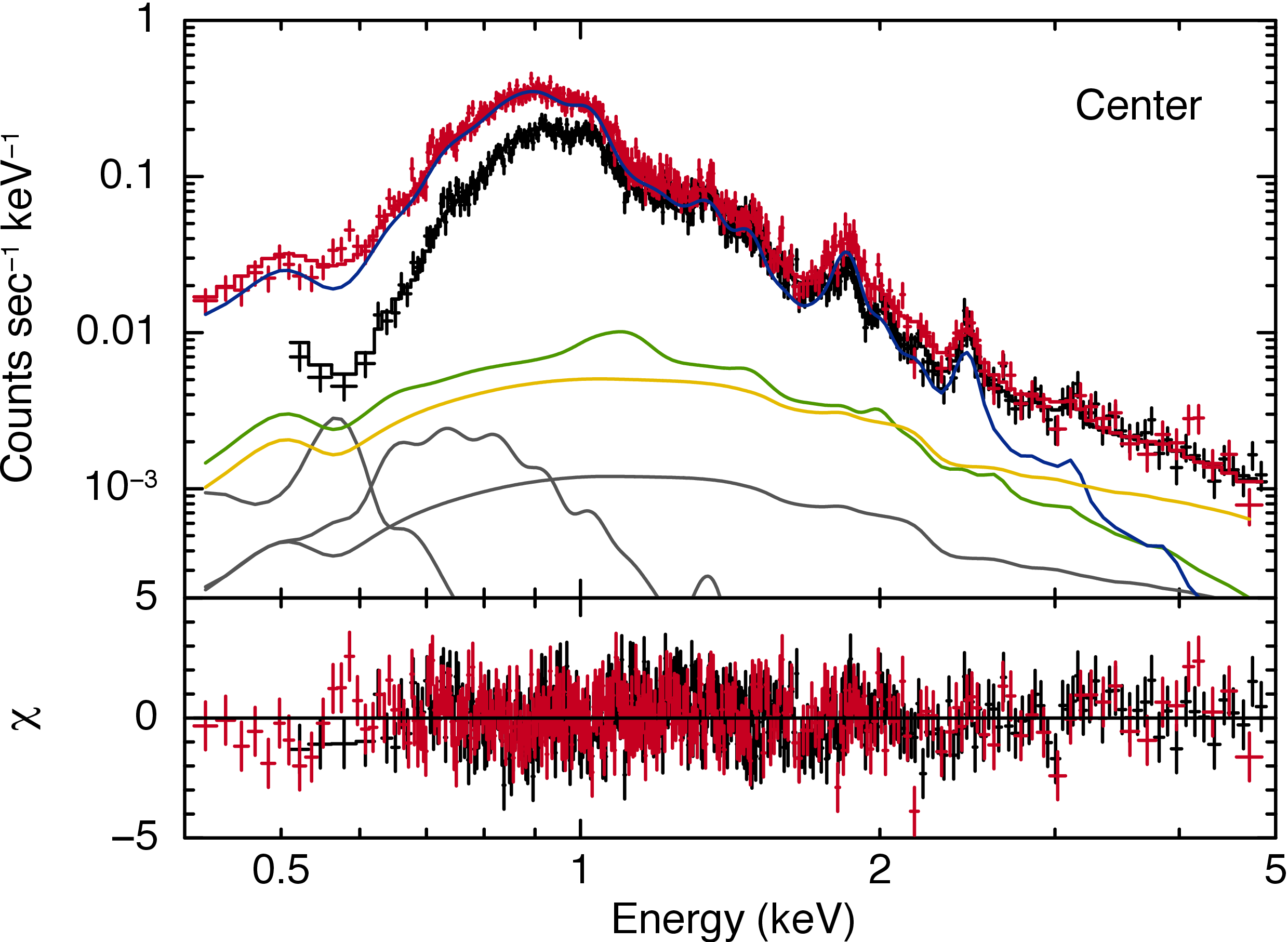}
\hspace{5mm}
\includegraphics[width=80mm]{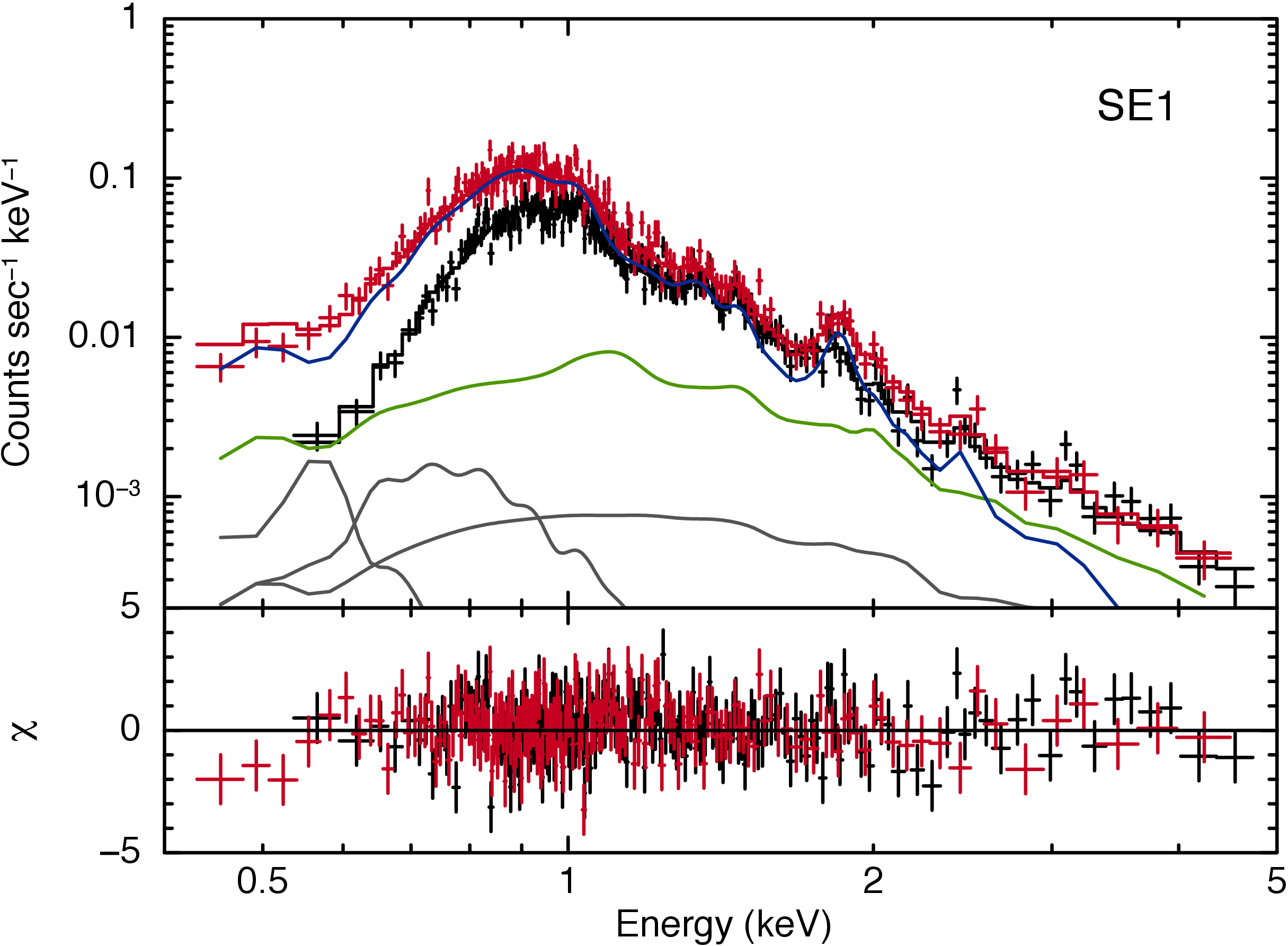}\\
\vspace{5mm}
\includegraphics[width=80mm]{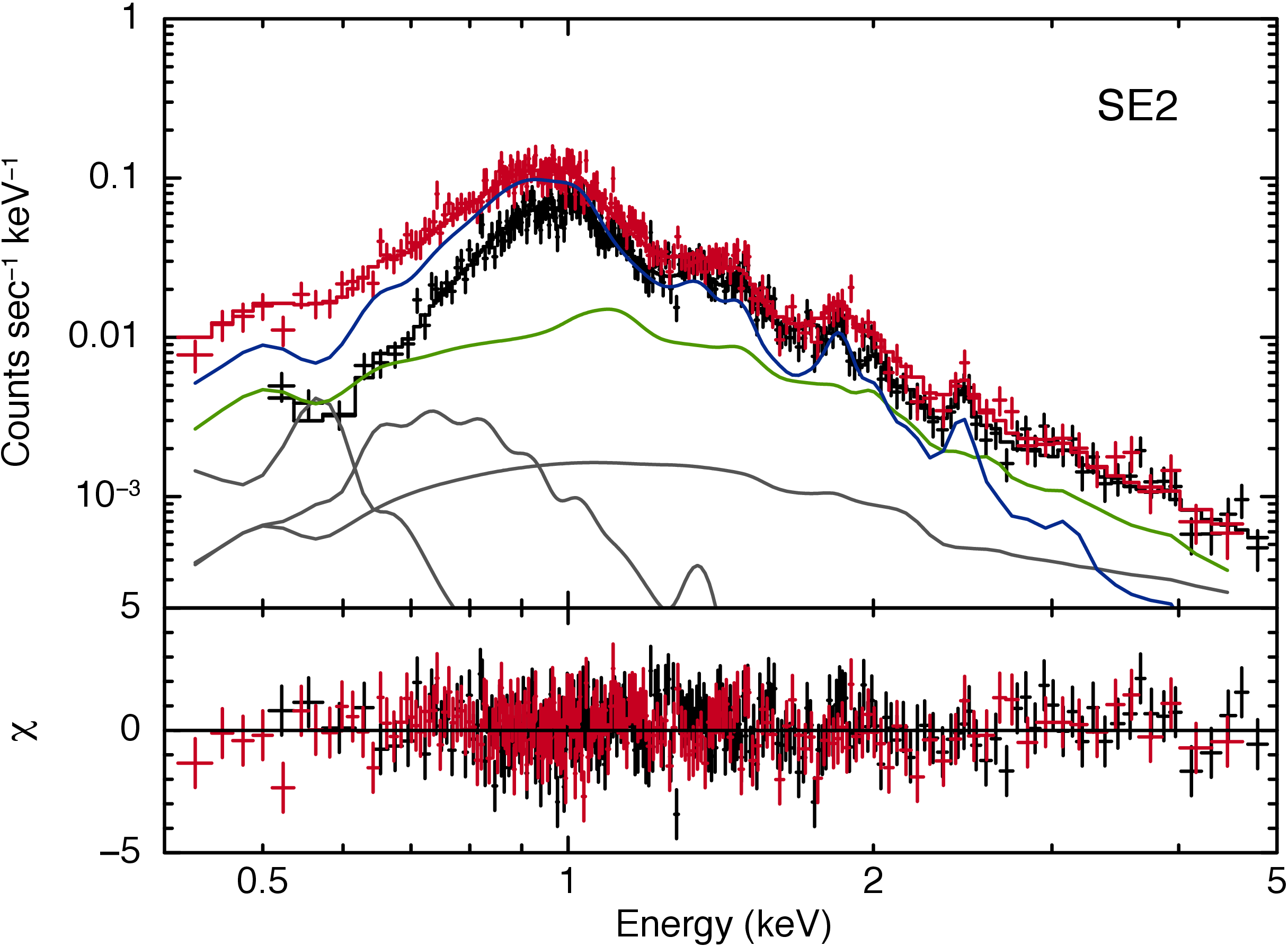}
\hspace{5mm}
\includegraphics[width=80mm]{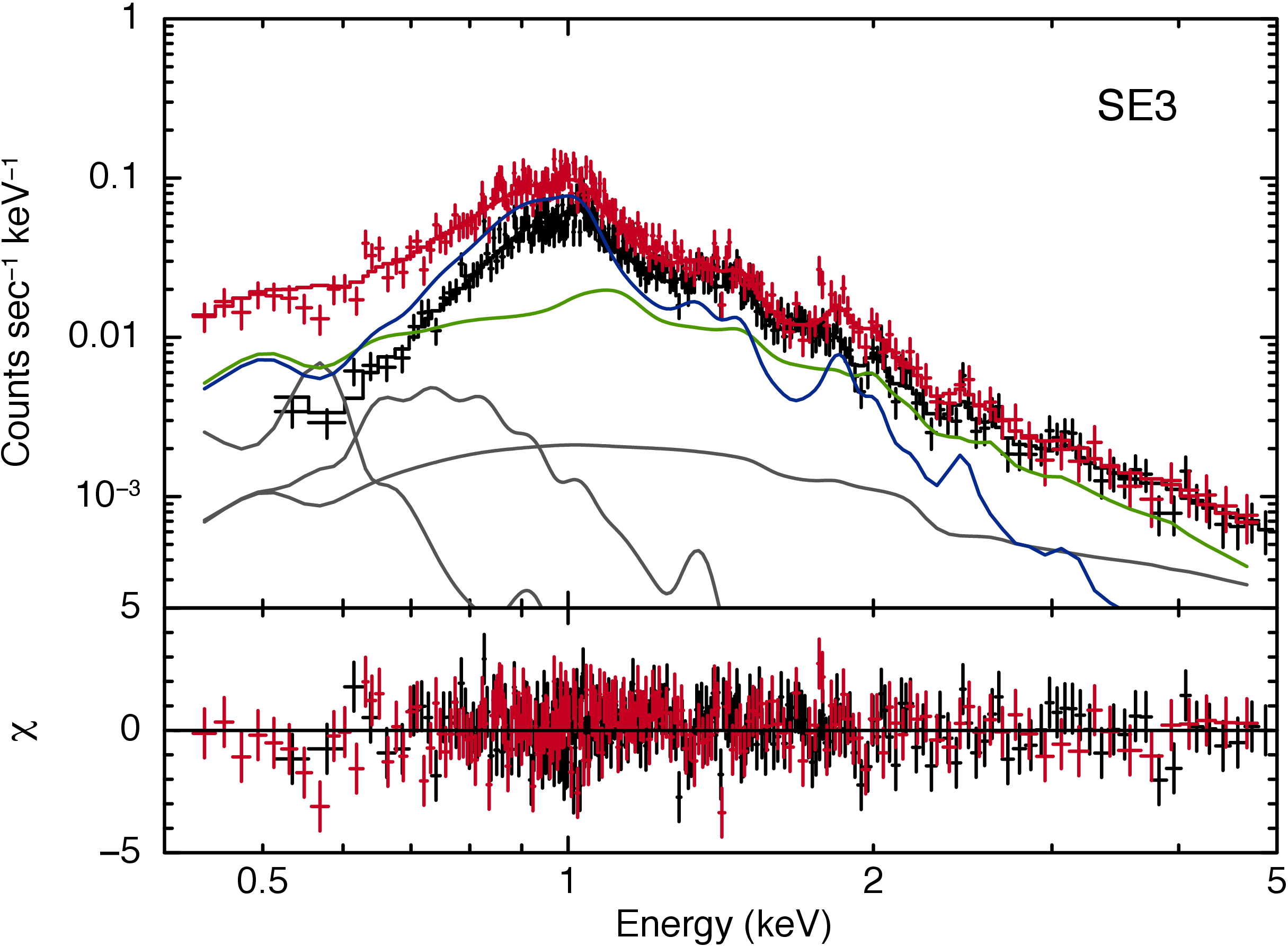}\\
\vspace{5mm}
\includegraphics[width=80mm]{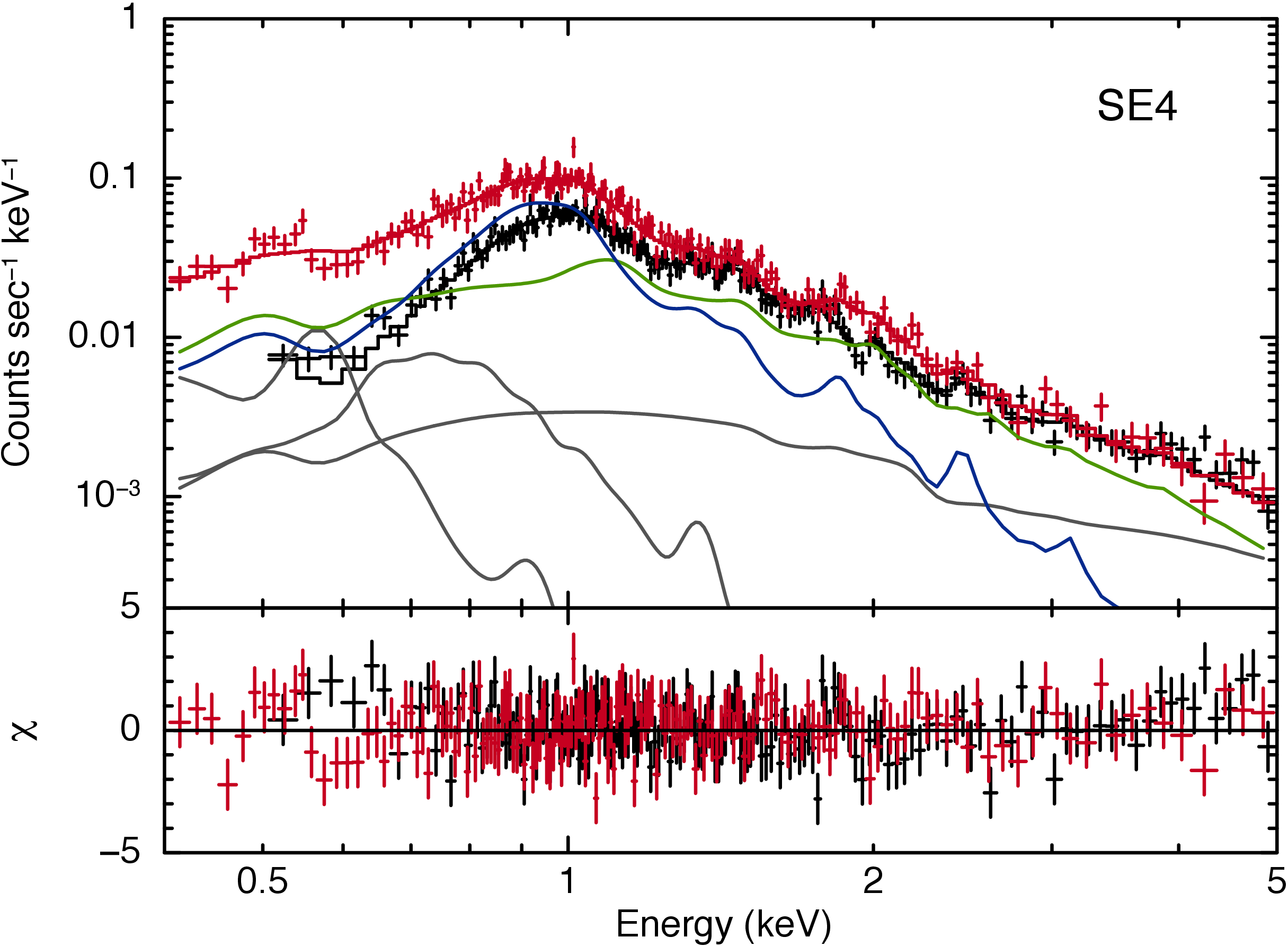}
\hspace{5mm}
\includegraphics[width=80mm]{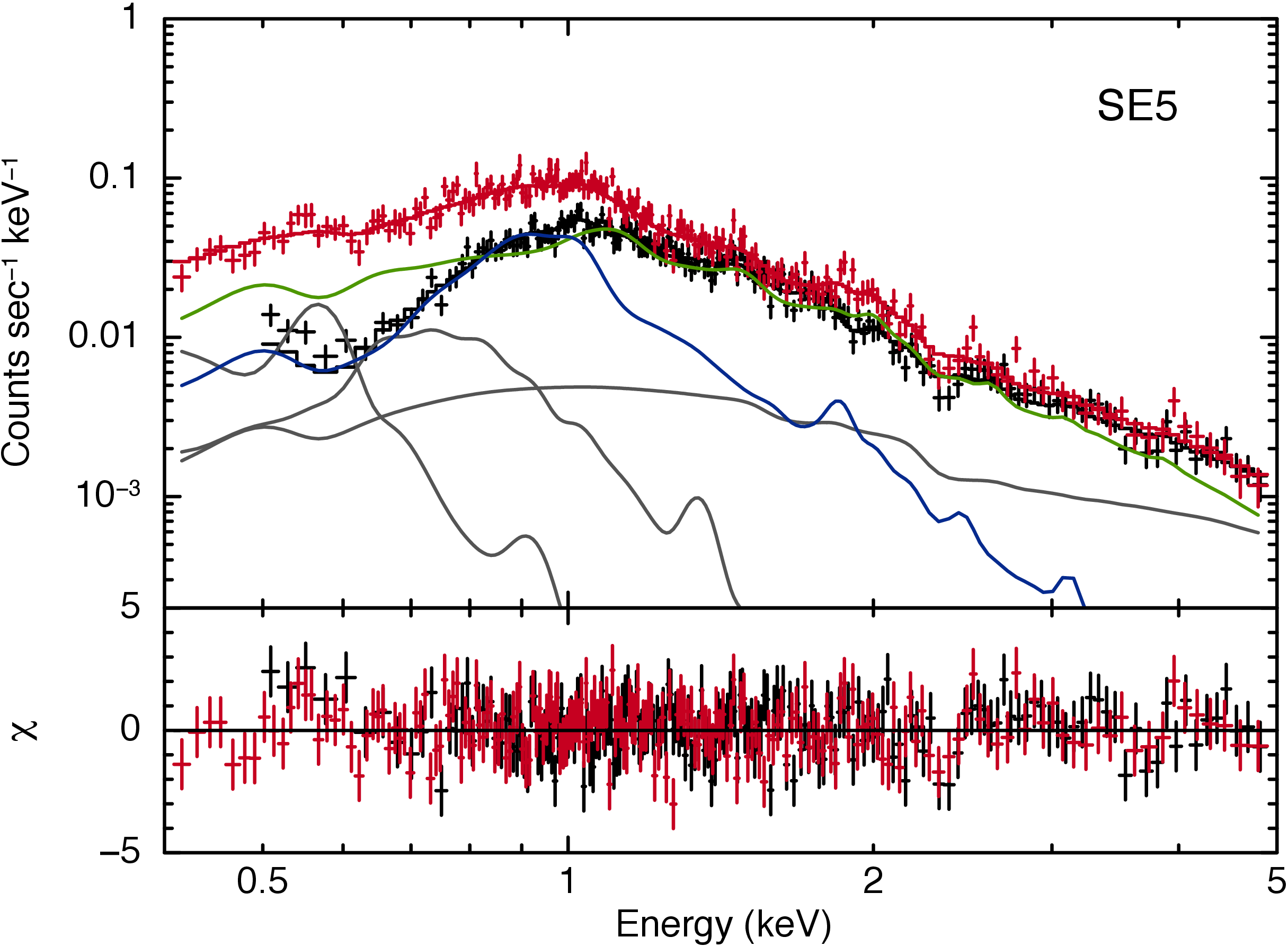}
\end{center}
\caption{Spectra of the center and SE1--6, and the best-fit models of
the $1T$ fit.  The red and black crosses show the data points of BI and
FI CCD data, and the red and black solid curves are the best fit models
for them.  The blue, green, yellow, and gray curves are the vAPEC
component, the ICM, the LMXBs, and the backgrounds/foreground components
(CXB, LHB, MWH), respectively.  Only the components for the BI model are
shown.}
\label{fig:fit_1T}
\end{figure*}

%%%%%%%%%%%%%%%%%%%%%%%%%%%%%%%%%%%%%%%%%%%%%%%%%%%%%%%%%%%%%%%%%%%%%%%%%
\begin{table*}
\tbl{Best-fit spectral parameters for EX 1--3 regions obtained
from the $1T$ fit and the $2T$ fit.}{
\begin{tabular}{ccccccccc}
\hline 
& & & \multicolumn{2}{c}{Region EX1} & \multicolumn{2}{c}{Region EX2} & \multicolumn{2}{c}{Region EX3}\\
Component & Parameter & Unit & $1T$ model & $2T$ model & $1T$ model & $2T$ model & $1T$ model & $2T$ model\\
\hline
vAPEC1 & $kT$  & (keV)   & 0.829$_{-0.008}^{+0.007}$ & 0.954$_{-0.021}^{+0.025}$ & 0.968$_{-0.012}^{+0.011}$ & 0.889$_{-0.038}^{+0.046}$ & 0.819$_{-0.021}^{+0.020}$ & 0.944$_{-0.055}^{+0.504}$\\
& $Z_{\rm O}$  & (solar) & 0.594$_{-0.166}^{+0.192}$ & 0.697$_{-0.185}^{+0.219}$ & 0.530$_{-0.322}^{+0.427}$ & 0.681$_{-0.353}^{+0.501}$ & 0.238$_{-0.184}^{+0.223}$ & 0.263$_{-0.189}^{+0.241}$\\
& $Z_{\rm Ne}$ & (solar) & 2.652$_{-0.370}^{+0.441}$ & 1.639$_{-0.364}^{+0.414}$ & 2.749$_{-0.796}^{+1.103}$ & 1.390$_{-0.327}^{+0.514}$ & 0.967$_{-0.305}^{+0.373}$ & 0.577$_{-0.372}^{+0.408}$\\
& $Z_{\rm Mg}$ & (solar) & 0.736$_{-0.103}^{+0.122}$ & 0.819$_{-0.119}^{+0.143}$ & 1.137$_{-0.257}^{+0.372}$ & 1.390$_{-0.327}^{+0.514}$ & 0.267$_{-0.096}^{+0.113}$ & 0.291$_{-0.103}^{+0.122}$\\
& $Z_{\rm Si}$ & (solar) & 0.531$_{-0.070}^{+0.080}$ & 0.653$_{-0.085}^{+0.099}$ & 0.909$_{-0.178}^{+0.252}$ & 1.076$_{-0.227}^{+0.350}$ & 0.278$_{-0.082}^{+0.091}$ & 0.308$_{-0.087}^{+0.097}$\\
& $Z_{\rm S}$  & (solar) & 0.589$_{-0.137}^{+0.145}$ & 0.619$_{-0.136}^{+0.146}$ & 1.123$_{-0.295}^{+0.373}$ & 1.214$_{-0.321}^{+0.449}$ & 0.350$_{-0.222}^{+0.226}$ & 0.361$_{-0.208}^{+0.211}$\\
& $Z_{\rm Fe}$ & (solar) & 0.547$_{-0.056}^{+0.068}$ & 0.724$_{-0.082}^{+0.101}$ & 0.852$_{-0.163}^{+0.249}$ & 1.024$_{-0.221}^{+0.359}$ & 0.211$_{-0.034}^{+0.045}$ & 0.260$_{-0.049}^{+0.064}$\\
& Norm &($\times 10^{-2}$) & 6.823$_{-0.680}^{+0.682}$ & 5.128$_{-0.704}^{+0.760}$ & 1.662$_{-0.350}^{+0.352}$ & 1.056$_{-0.356}^{+0.445}$ & 1.768$_{-0.254}^{+0.255}$ & 1.343$_{-0.768}^{+0.340}$\\
%\hline
vAPEC2 & $kT$&(keV) & -- & 0.606$_{-0.078}^{+0.061}$ & -- & 1.461$_{-0.182}^{+0.339}$ & -- & 0.630$_{-0.201}^{+0.818}$\\
& Norm &($\times 10^{-2}$) &--&1.392$_{-0.401}^{+0.526}$ &--&0.709$_{-0.223}^{+0.287}$ &--&0.442$_{-0.225}^{+1.144}$\\
APEC (ICM) & $kT$ & (keV) & \multicolumn{6}{c}{2.1 (fixed)}\\
 & $Z$ & (solar) & \multicolumn{6}{c}{0.27 (fixed)}\\
 & Norm &($\times 10^{-2}$) & 3.582$_{-0.234}^{+0.233}$&3.314$_{-0.274}^{+0.272}$ & 3.148$_{-0.174}^{+0.174}$&2.662$_{-0.607}^{+0.333}$ & 2.793$_{-0.100}^{+0.099}$&2.708$_{-0.128}^{+0.130}$\\
%\hline
-- & $\chi^2/{\rm d.o.f.}$ &&959.30/729&854.51/727& 979.52/838&928.26/836&1039.13/852&1029.93/850 \\
\hline
\end{tabular}}
\label{tab:spec_param_EX}
\begin{tabnote}
In all the cases, APEC models for LHB ($kT=0.11$~keV, $Z=1Z_\odot$,
$\textrm{Norm}=9.6\times10^{-3}$) and MWH ($kT=0.3$~keV, $Z=1Z_\odot$,
$\textrm{Norm}=1.9\times10^{-3}$), and a power law model for CXB
($\Gamma=1.4$, $\textrm{Norm}=1.063\times10^{-3}$) were included. The
normalizations of the APEC components are in units of
$\frac{10^{-14}}{4\pi[D_A(1+z)]^2}\int n_en_H dV$ per $400\pi$
arcmin$^2$, where $D_A$ is the angular diameter distance to the source
(cm), $n_e$ and $n_H$ are the electron and hydrogen number densities
(cm$^{-3}$). The normalizations of the power law are in units of
photons\,keV$^{-1}$\,cm$^{-2}$\,s$^{-2}$ at 1~keV per
$400\pi$~arcmin$^2$.
\end{tabnote}
\end{table*}

\begin{figure*}
\begin{center}
\includegraphics[width=80mm]{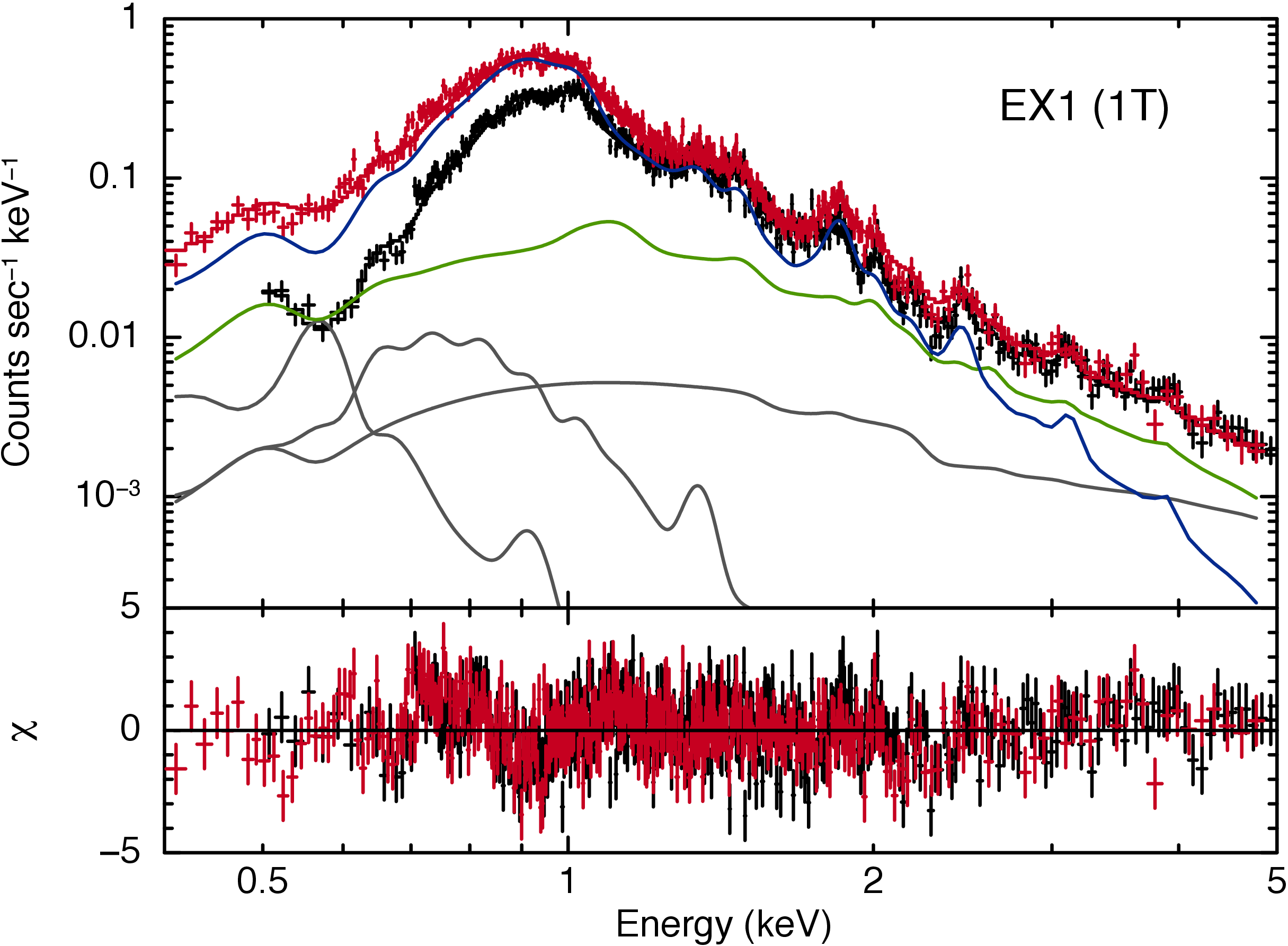}
\hspace{5mm}
\includegraphics[width=80mm]{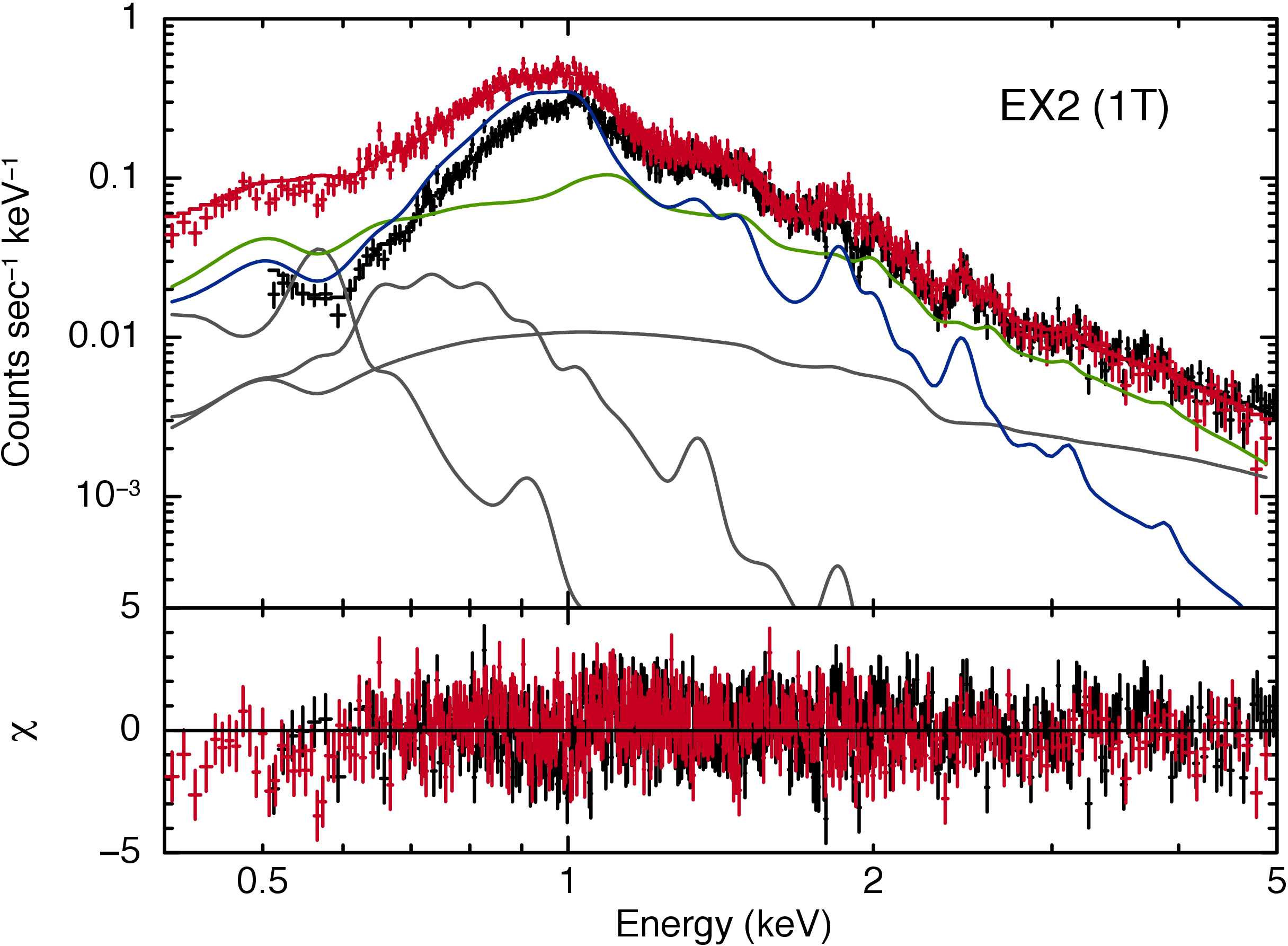}
\hspace{5mm}
\includegraphics[width=80mm]{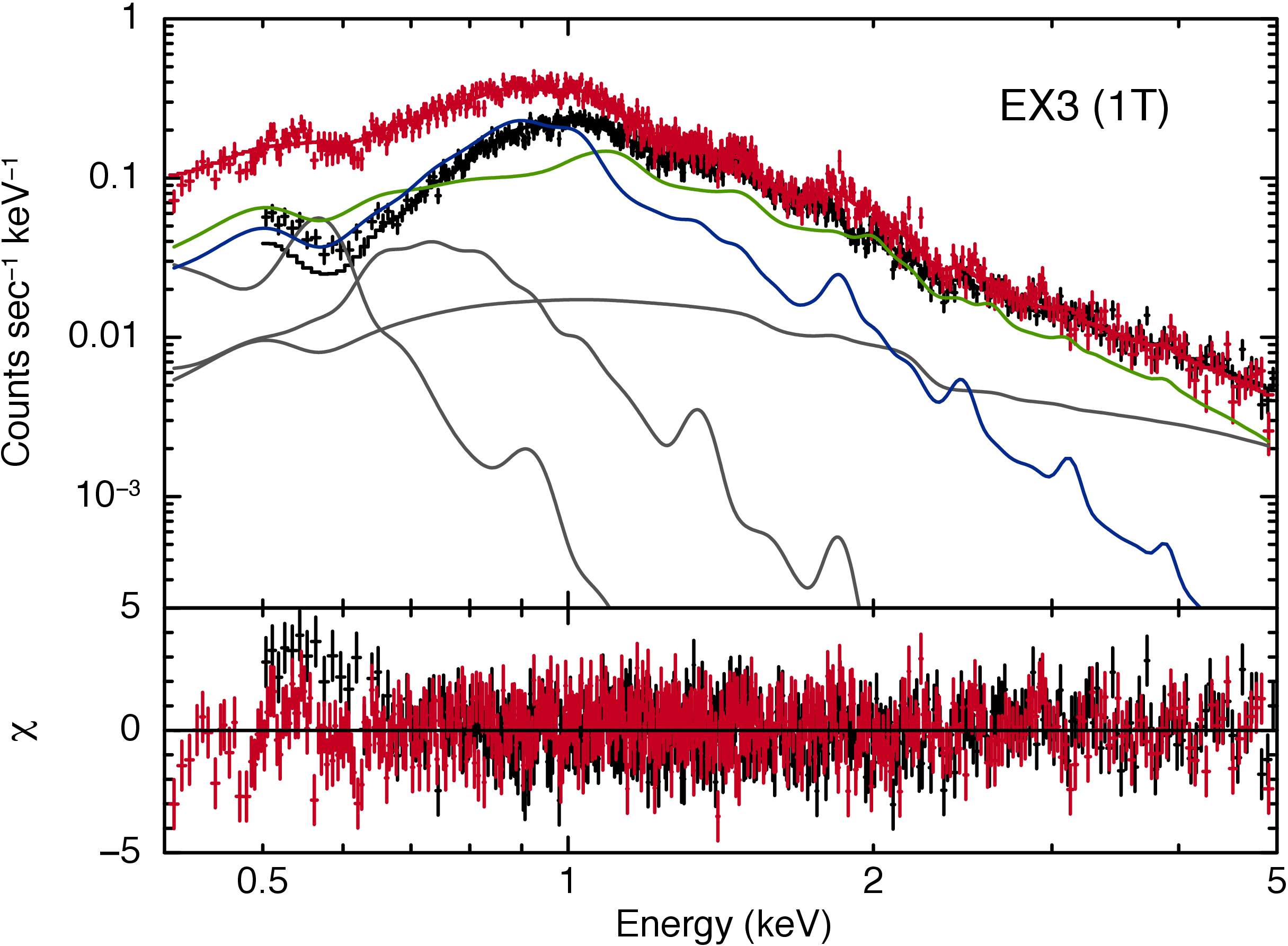}
\end{center}
\caption{Spectra of the center and EX1--3, and the best-fit models of
the $1T$ fit. The red and black crosses show the data points of BI and
FI CCD data, and the red and black solid curves are the best fit models
for them.  The blue, green, and gray curves are the vAPEC component, the
ICM, and the backgrounds/foreground components (CXB, LHB, MWH),
respectively.  Only the components for the BI model are shown.}
\label{fig:fit_EX}
\end{figure*}

We fitted the spectra with a single-temperature ($1T$) model and a
two-temperature ($2T$) model, represented by one or two vAPEC model(s),
which is an APEC model with variable abundances, in addition to the
background and foreground components described in the previous
section. An additional power law was added to the center region, to
represent the contribution of unresolved low mass X-ray binaries.  The
photon index was fixed at 1.5 (e.g., \cite{Sarazin2003}). To avoid the
normalizations of the galactic components varying region by region, we
first fitted the center and SE1--6 regions simultaneously to determine
the normalizations of the galactic components, and fixed them at the
values obtained in the simultaneous fitting.  The fitting results of the
$1T$ model and the $2T$ model are shown in table~\ref{tab:SE_1T} for the
center and SE1--5 regions, and in table~\ref{tab:spec_param_EX}
for EX1--3 regions. The spectra and the best-fit models of the $1T$ fit
are shown in figure~\ref{fig:fit_1T} for the center and SE1--5, and in
figure~\ref{fig:fit_EX} for EX1--3.
Since there was an uncertainty in the normalizations of the LHB and the
MWH as described in section~\ref{sec:Outer_regions}, we investigated
how the results of the
$1T$ fit were affected if these normalizations were fixed at the numbers
obtained by \citet{Simionescu15}. When we fixed the normalization of
the MWH at $1.9\times10^{-4}$, i.e., 1/10 of the number shown in table 4
and 5, the best-fit parameters were unchanged within a statistical error range
even in SE4 and SE5. When we fixed the normalization of the LHB
at $0.16\times10^{-3}$, i.e., 1.6\% of the number shown in table 4 and
5, the temperature of the vAPEC component of SE5 decreased, the Fe
abundances of SE4 and SE5 decreased, and the normalizations of SE4 and
SE5 increased, while other parameters were unchanged within 
a statistical error range.
In these cases, however, reduced $\chi^2$ increased
by 0.05 for SE4 and 0.11 for SE5. When another APEC component was added,
the temperature became close to that of the LHB. Therefore, a larger
normalization of the LHB or equivalent was needed to represent the
Suzaku spectra.

%%%%%%%%%%%%%%%%%%%%%%%%%%%%%%%%%%%%%%%%%%%%%%%%%%%%%%%%%%%%%%%%%%%%%%%%%
\subsubsection{Temperatures}

\begin{figure}
\begin{center}
\includegraphics[width=80mm]{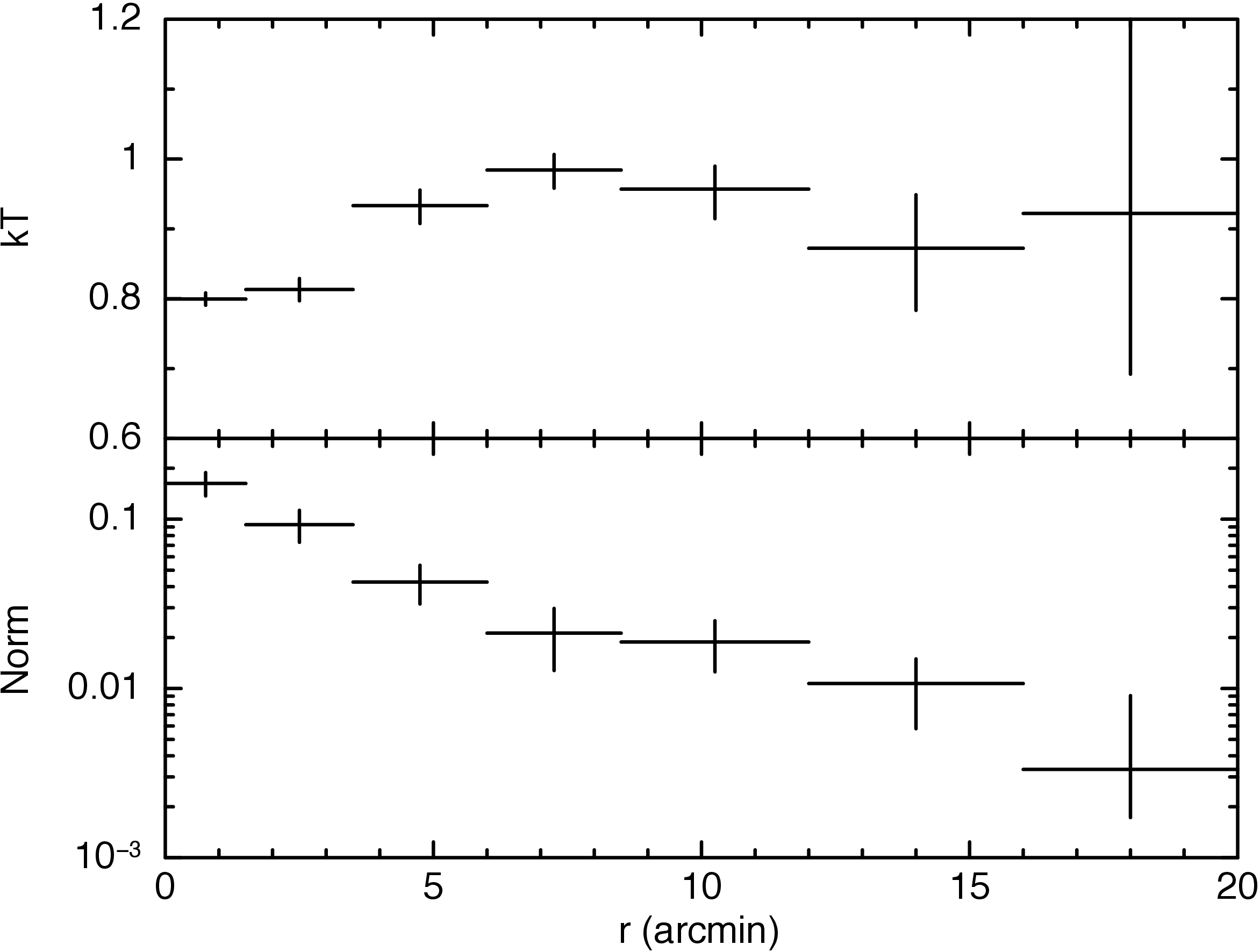}
\end{center}
\caption{Radial profiles of the temperature and the normalization of the
vAPEC component of the $1T$ model of the center and SE1--6 regions.}
\label{fig:1T_radial}
\end{figure}

The temperatures of the $1T$ fit were about 0.8--1.0~keV. As shown in
the upper panel of figure~\ref{fig:1T_radial}, the temperatures of the
inner regions (center and SE1) were lower, while they were almost
constant or slightly decreasing toward the outer regions (SE2--6).  When
the $1T$ fit and the $2T$ fit were compared, the $F$-test probabilities
were $8.5\times10^{-4}$, $9.0\times10^{-4}$, $2.1\times10^{-3}$,
$1.5\times10^{-3}$, $6.5\times10^{-3}$, and 0.37, for the center and the
SE1--5 regions, respectively.  Thus, the improvement of the fit was
reasonable except for SE5.

In the center region, the temperature of the main component was
$0.88^{+0.03}_{-0.04}$~keV while that of the second component was
$\sim0.6$~keV, when the $2T$ model was employed. The normalization of
the second component was about 0.3 times that of the first
component. When the ICM temperature was made free in the $1T$ fit, it
resulted in $\sim0.6$~keV, rather than staying around 2~keV. Thus, the
spectral data preferred the existence of a cold
component. \citet{Matsushita01} showed that the temperature of the
central region ($<\timeform{2.6'}$) was 0.69~keV, while
\citet{M86_Chandra} reported existence of cold clumps around the
core. \citet{M86_XMM2} pointed the presence of $\sim0.6$--0.7~keV gas
between M86 and NGC~4438. Our result was qualitatively consistent with
them.

For SE1, the temperature of the main component was almost unchanged
while the second temperature was $1.82^{+0.22}_{-0.32}$~keV. The
normalization of the ICM, however, became 0, which was unrealistic. If
the ICM temperature was made free in the $1T$ fit, it became
$1.56^{+0.26}_{-0.17}$~keV, and the abundances increased by
$\sim0.1Z_\odot$.  The $F$-test probability was $7.7\times10^{-3}$, and
hence, this improvement was reasonable. The results suggest that the
$1T$ model is enough, but the ICM temperature could be as low as
$\sim1.6$~keV in SE1. Note that it is close to that of region S shown in
table~\ref{tab:outer_region_params}.

The second temperatures of SE2 and SE3 were 1.3~keV and 1.7~keV,
respectively, suggesting the ICM temperature was lower than what we
assumed like SE1.  When the ICM temperature was made free in the $1T$
fitting, they became $1.50^{+0.32}_{-0.17}$~keV and
$1.71^{+0.12}_{-0.13}$~keV, respectively.  In these cases, however, the
abundances of the main component became unphysically large, and thus,
the results were considered unacceptable.  This is probably because the
room for the continuum became smaller, as the ICM temperature became
lower.  The results may suggest that the ICM temperature is located
between them, but it was not possible to further constrain them.
For SE4, the second temperature was too high to constrain.

As a conclusion, the $2T$ model is better for the center, while the $1T$
model is enough for SE1 but the ICM temperature could be as low as
1.6~keV. For SE2 and SE3, the ICM temperature may also be lower than
2.1~keV, while that of SE4 may be slightly higher.

Note that the temperatures of EX1--3 showed a similar characteristics.
When the $1T$ fit and $2T$ fit were compared, the $F$-test probabilities
were $6\times10^{-19}$, $2\times10^{-10}$, and 0.023, for EX1, 2, 3,
respectively. Thus, the improvement of the fit was reasonable for EX1
and 2.  The second temperature of EX1 was $0.61^{+0.06}_{-0.08}$~keV,
which may be the cooler component either located at the center or in the
region between M86 and NGC~4438. The results of EX2 may indicate that
the ICM temperature is lower than what was assumed.

%%%%%%%%%%%%%%%%%%%%%%%%%%%%%%%%%%%%%%%%%%%%%%%%%%%%%%%%%%%%%%%%%%%%%%%%%
\subsubsection{Normalizations, density and mass of the core}
\label{sec:core_halo_norm}

The normalizations of the $1T$ fit were plotted as a function of radius
from the center, in the lower panel of figure~\ref{fig:1T_radial}. When
the profile was fitted with a $\beta$-model, the best-fit parameters
were $\beta=0.42^{+0.17}_{-0.09}$, $\theta_0=2.0^{+2.1}_{-1.3}$, and
$S_0=0.18^{+0.09}_{-0.05}$. $\beta$ and $\theta_0$ were consistent with
those of sector 4 shown in table~\ref{tab:sb_fit} within an error range.

From the normalization of the APEC model, the emission
measure can be obtained.  Assuming the center region as a uniform sphere
of \timeform{1.5'} (7.2~kpc) radius, the number density of hydrogen $n_H$ became
$\sim7.1\times10^{-3}$~cm$^{-3}$, and the mass of the sphere $M$ became
$\sim3.9\times10^8~M_\odot$, where mean molecular weight of hydrogen was
assumed to be 1.4.  When the normalization of the $2T$ model was
adopted, the results were unchanged ($n_H\sim6.2\times10^{-3}$~cm$^{-3}$
and $M\sim3.4\times10^8M_\odot$). Note that they are consistent with
what was obtained by \citet{M86_Chandra} ($n_{\rm
core}\approx6.2\times10^{-3}$~cm$^{-3}$ and $M_{\rm
core}\approx7.4\times10^8~M_\odot$ within a sphere of radius 9.6~kpc).

%%%%%%%%%%%%%%%%%%%%%%%%%%%%%%%%%%%%%%%%%%%%%%%%%%%%%%%%%%%%%%%%%%%%%%%%%
\subsubsection{Abundances}

\begin{figure*}
\begin{center}
\includegraphics[width=80mm]{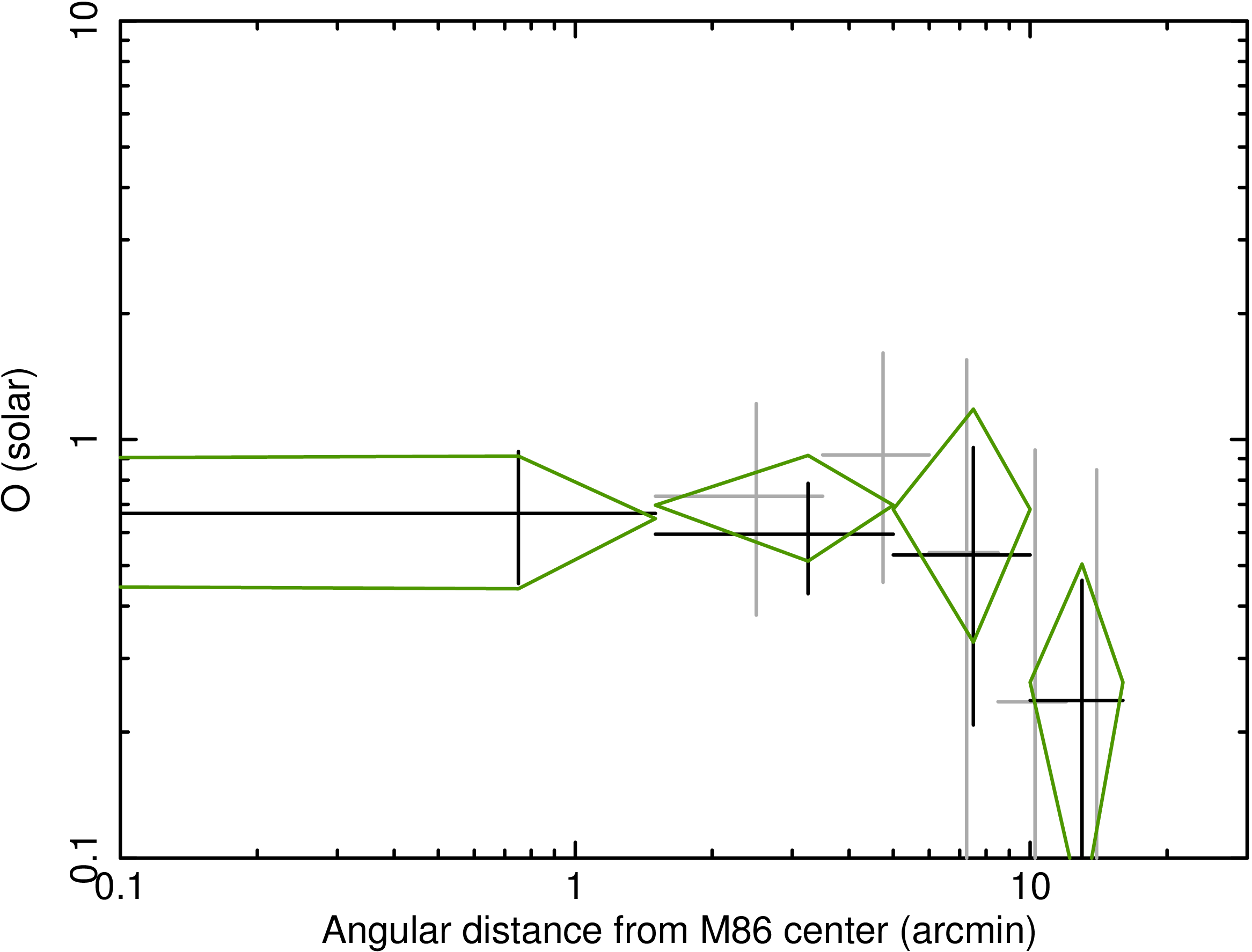}
\hspace{5mm}
\includegraphics[width=80mm]{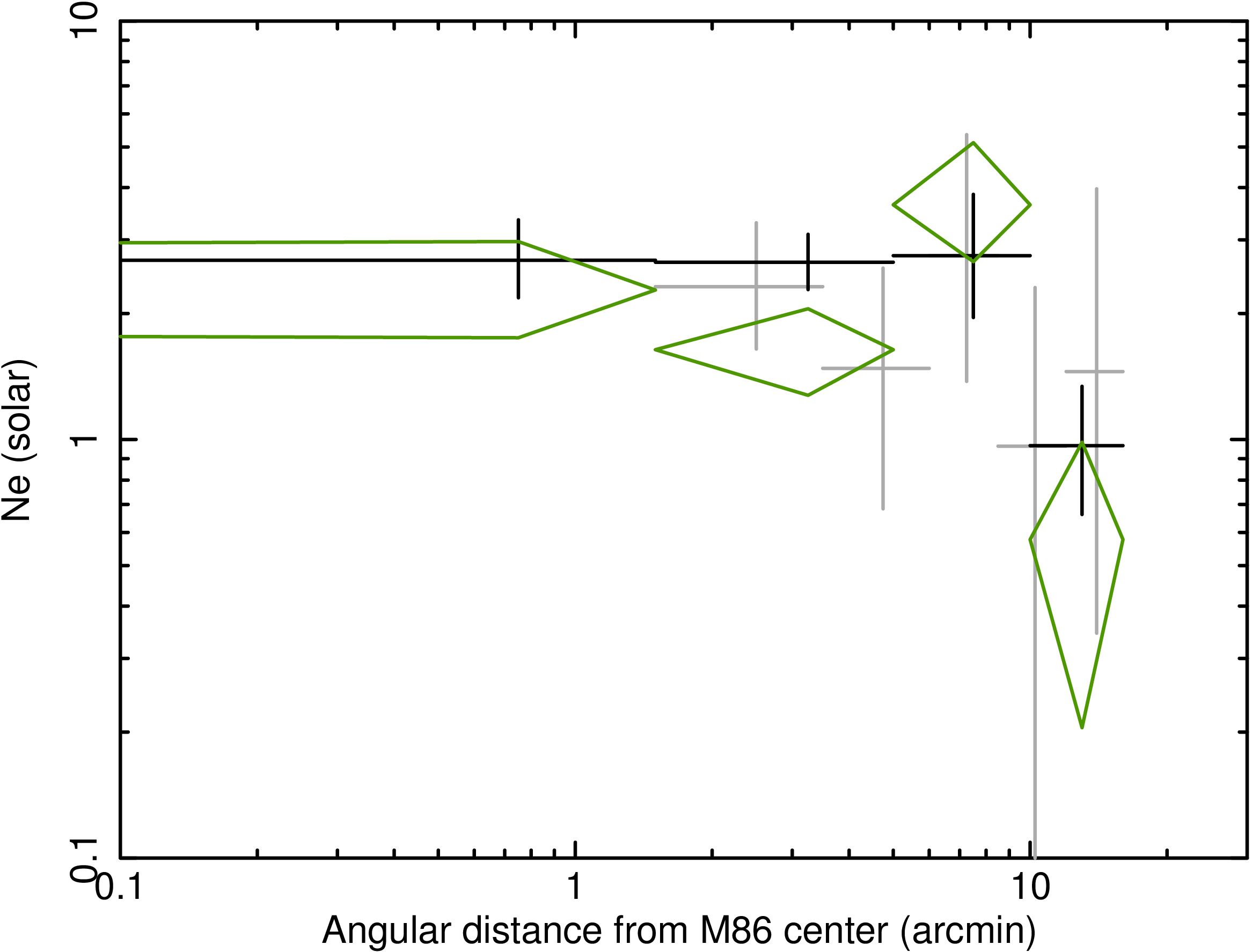}\\
\includegraphics[width=80mm]{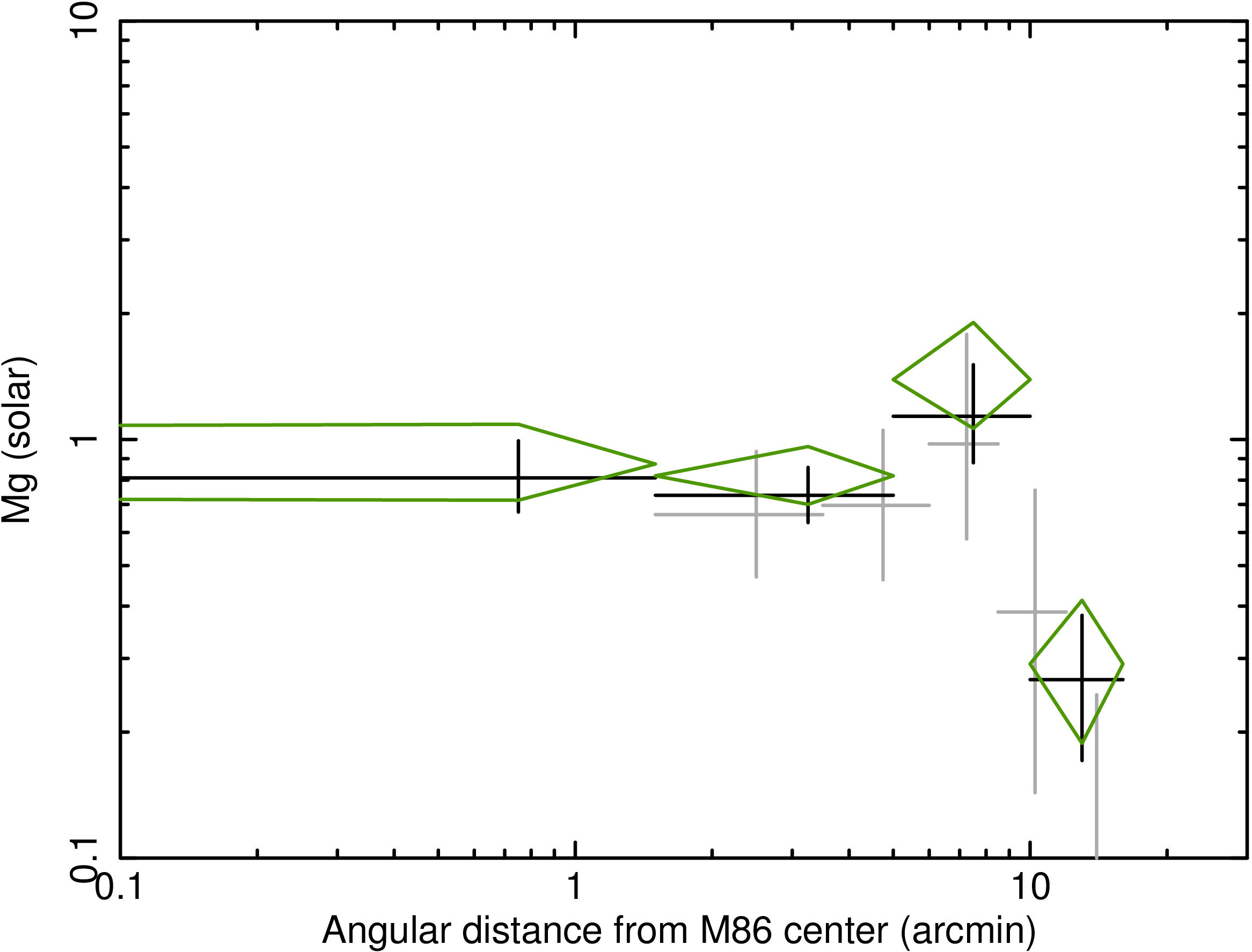}
\hspace{5mm}
\includegraphics[width=80mm]{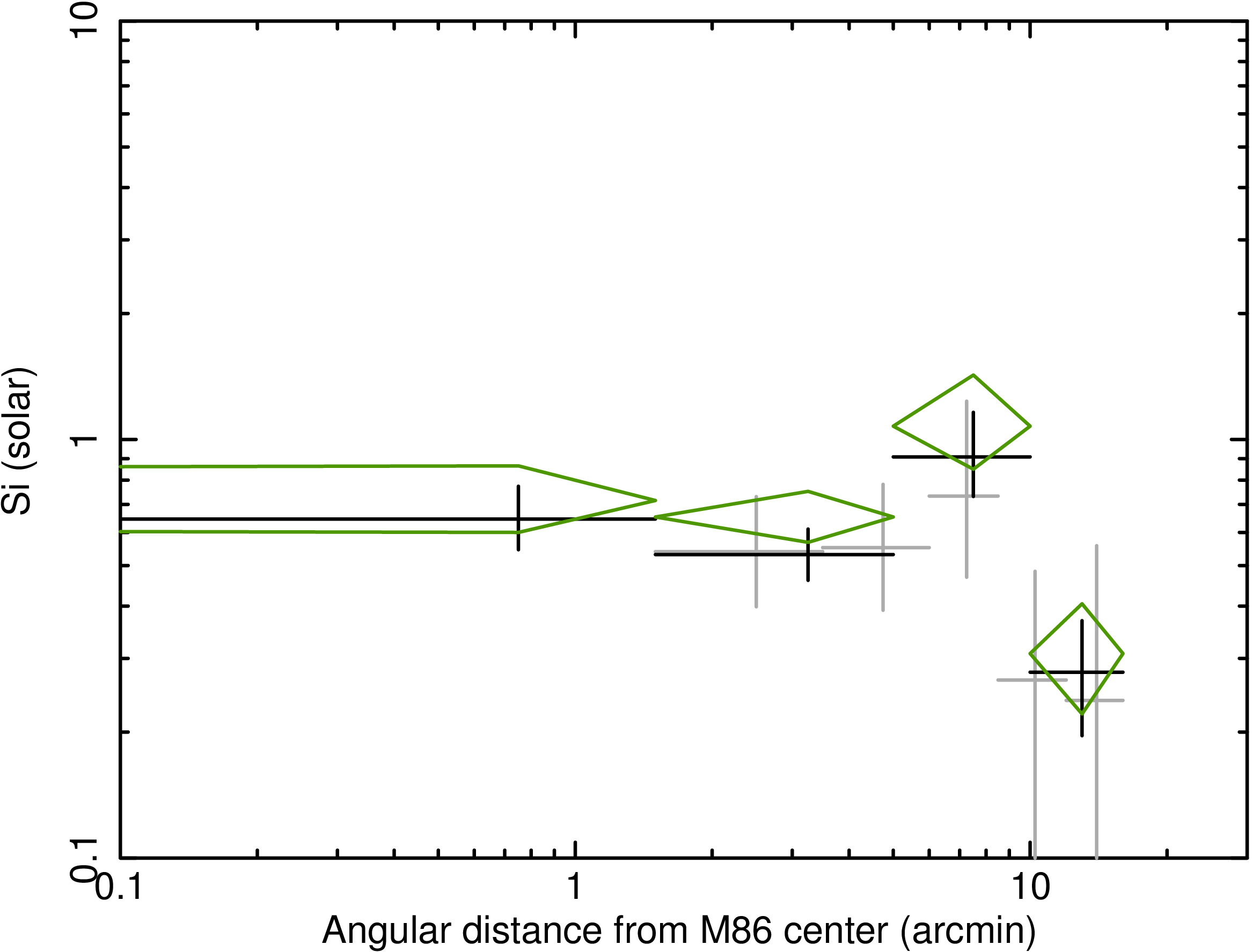}\\
\includegraphics[width=80mm]{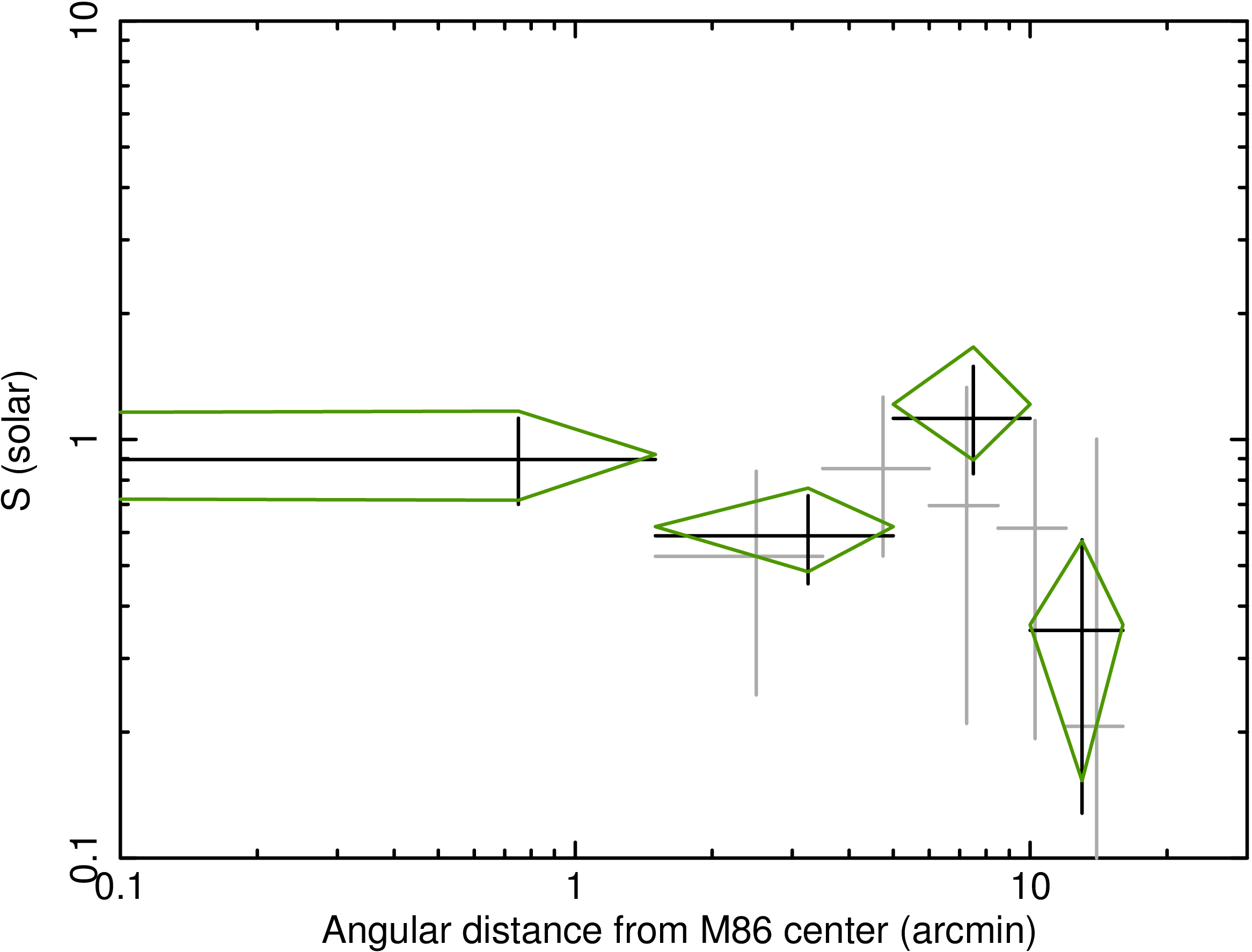}
\hspace{5mm}
\includegraphics[width=80mm]{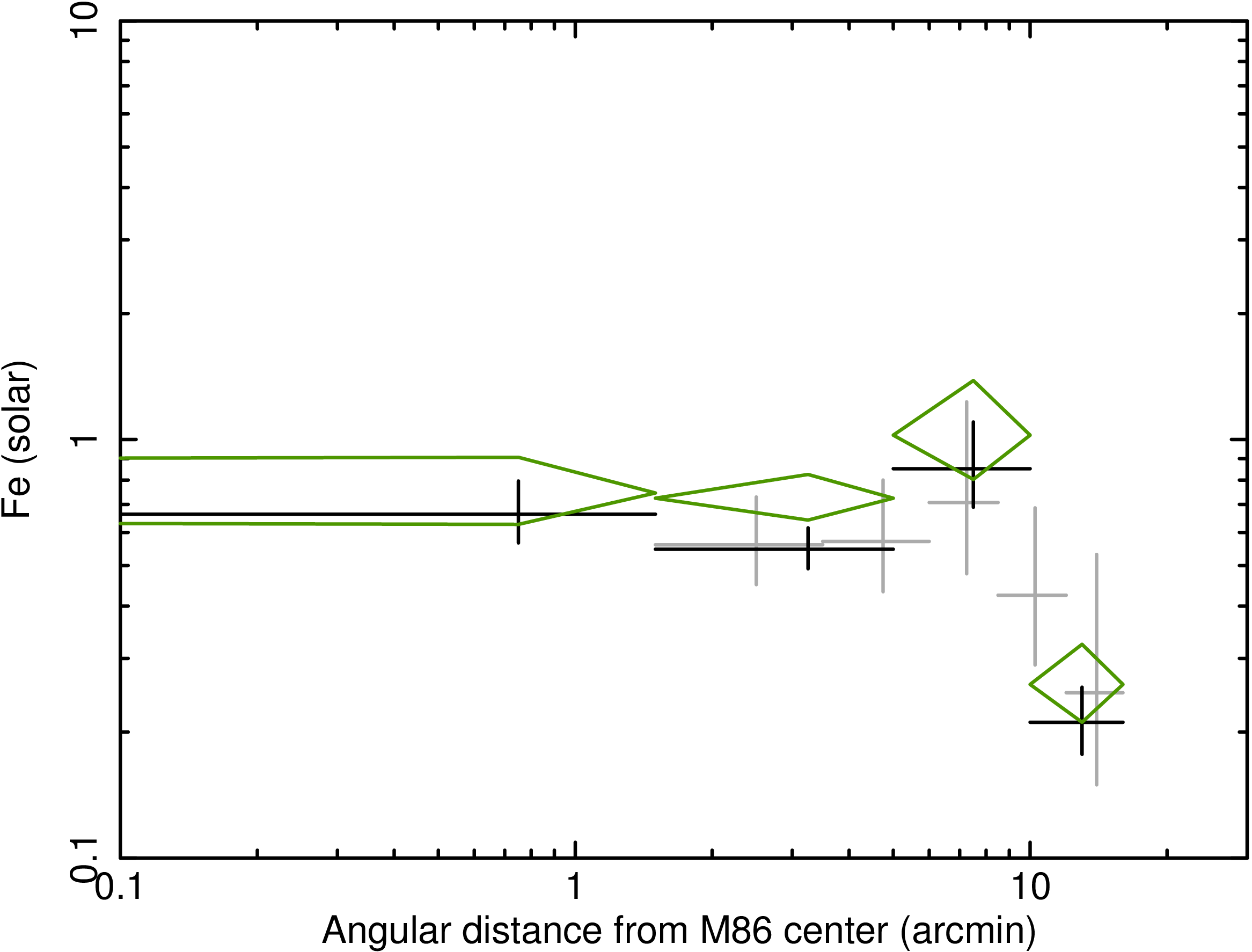}\\
\end{center}
\caption{Radial distributions of the abundances of O, Ne, Mg, Si, S, and
Fe, as a function of angular distance from the M86 center.  Black
crosses and green squares are $1T$ fit and $2T$ fit results of the
center and EX1--3 regions, respectively.  $1T$ fit results of SE1--5 are
also shown with gray crosses.}
\label{fig:abundance_distribution}
\end{figure*}

Figure~\ref{fig:abundance_distribution} shows the radial distributions
of the abundances of O, Ne, Mg, Si, S, and Fe of the center and EX1--3
regions.  $1T$ fit results of SE1--5 regions are also shown. The
distributions were consistent with each other, although the errors of
SE1--5 were large.  Within $r\lesssim\timeform{10'}$, the abundances
were relatively large and roughly constant, while outside
$r\gtrsim\timeform{10'}$ the abundances became significantly smaller
(0.2--$0.3Z_\odot$).

\begin{figure}
\begin{center}
\includegraphics[width=80mm]{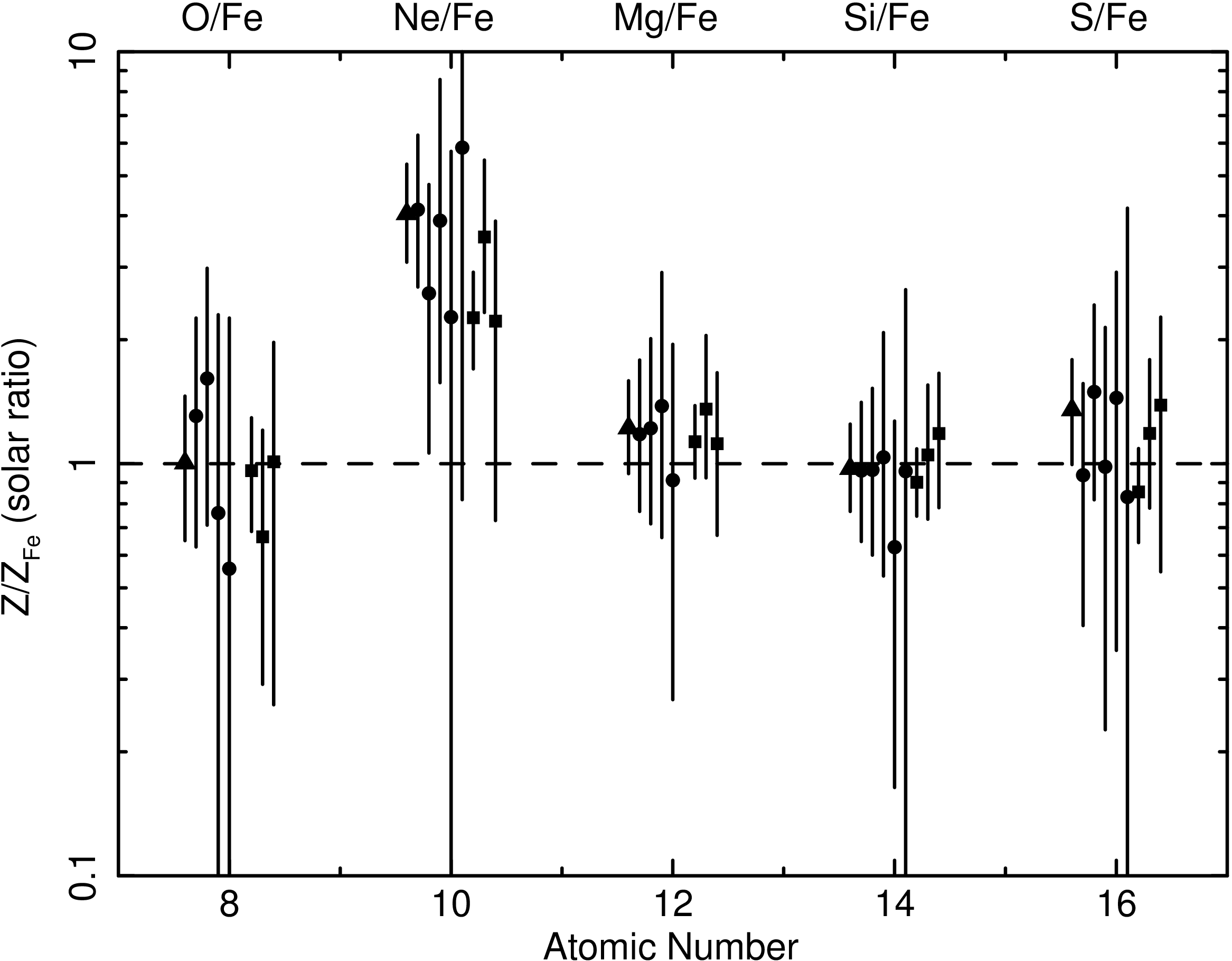}
\end{center}
\caption{
Abundance ratios of O, Ne, Mg, Si, and S with respect to Fe. Triangles,
circles, and squares represent the center, SE1--5, and EX1--3,
respectively.}
\label{fig:abundance_ratio}
\end{figure}

Figure~\ref{fig:abundance_ratio} shows the abundance ratios of elements
with respect to Fe. O/Fe, Mg/Fe, Si/Fe, S/Fe are consistent with 1,
while Ne/Fe is 2--4. When the $2T$ model was adopted, the abundances
were generally slightly higher, by $\sim0.1Z_\odot$, but the
overabundance of Ne was unchanged.

%%%%%%%%%%%%%%%%%%%%%%%%%%%%%%%%%%%%%%%%%%%%%%%%%%%%%%%%%%%%%%%%%%%%%%%%%
\subsection{Plume and tail}

\begin{table*}
\tbl{Best-fit parameters of the plume and the tail regions.}{
\begin{tabular}{ccccccc}
\hline 
& & & \multicolumn{2}{c}{Plume} & \multicolumn{2}{c}{Tail} \\
Component & Parameter & Unit & Case 1 & Case 2 & Case 1 & Case 2 \\
\hline 
vAPEC & $kT$&(keV) & 0.861$_{-0.007}^{+0.010}$ & 0.862$_{-0.007}^{+0.011}$ & 0.995$_{-0.011}^{+0.010}$ & 0.998$_{-0.011}^{+0.010}$ \\
& $Z_{\rm O}$ &(solar) & 0.751$_{-0.271}^{+0.333}$ & 0.574$_{-0.201}^{+0.229}$ & 1.290$_{-0.579}^{+0.926}$ & 1.012$_{-0.477}^{+0.806}$ \\
& $Z_{\rm Ne}$&(solar) & 2.668$_{-0.568}^{+0.717}$ & 2.068$_{-0.370}^{+0.515}$ & 3.637$_{-1.335}^{+2.039}$ & 2.781$_{-1.169}^{+1.981}$ \\
& $Z_{\rm Mg}$&(solar) & 1.218$_{-0.213}^{+0.271}$ & 0.999$_{-0.140}^{+0.188}$ & 1.695$_{-0.454}^{+0.773}$ & 1.371$_{-0.399}^{+0.740}$ \\
& $Z_{\rm Si}$&(solar) & 0.813$_{-0.131}^{+0.164}$ & 0.692$_{-0.091}^{+0.105}$ & 1.241$_{-0.298}^{+0.499}$ & 1.022$_{-0.264}^{+0.488}$ \\
& $Z_{\rm S}$ &(solar) & 0.990$_{-0.221}^{+0.251}$ & 0.861$_{-0.168}^{+0.179}$ & 1.316$_{-0.403}^{+0.584}$ & 1.095$_{-0.343}^{+0.551}$ \\
& $Z_{\rm Fe}$&(solar) & 0.906$_{-0.126}^{+0.167}$ & 0.715$_{-0.055}^{+0.121}$ & 1.377$_{-0.328}^{+0.583}$ & 1.084$_{-0.313}^{+0.603}$ \\
& Norm &($\times 10^{-2}$) & 10.230$_{-1.495}^{+1.511}$ & 12.856$_{-0.927}^{+0.521}$ & 4.186$_{-1.185}^{+1.190}$ & 5.340$_{-1.884}^{+2.007}$ \\
%\hline
APEC (ICM) & $kT$&(keV) & 2.1 (fixed) & $>3.2$ & 2.1 (fixed) & 2.728$_{-0.727}^{+6.861}$\\
 & Norm &($\times 10^{-2}$) & 2.095$_{-0.508}^{+0.497}$ & 0.862$_{-0.270}^{+0.180}$ & 3.006$_{-0.576}^{+0.575}$ & 2.113$_{-1.158}^{+1.301}$\\
%\hline
-- & $\chi^2/{\rm d.o.f.}$ & & 600.13/566 & 592.98/565 & 441.21/422 & 439.52/421\\
\hline
\end{tabular}}
\label{tab:plume_fit}
\begin{tabnote}
In all the cases, APEC models for LHB ($kT=0.11$~keV, $Z=1Z_\odot$,
$\textrm{Norm}=9.6\times10^{-3}$) and MWH ($kT=0.3$~keV, $Z=1Z_\odot$,
$\textrm{Norm}=1.9\times10^{-3}$), and a power law model for CXB
($\Gamma=1.4$, $\textrm{Norm}=1.063\times10^{-3}$) were included. The
normalizations of the APEC components are in units of
$\frac{10^{-14}}{4\pi[D_A(1+z)]^2}\int n_en_H dV$ per $400\pi$
arcmin$^2$, where $D_A$ is the angular diameter distance to the source
(cm), $n_e$ and $n_H$ are the electron and hydrogen number densities
(cm$^{-3}$). The normalizations of the power law are in units of
photons\,keV$^{-1}$\,cm$^{-2}$\,s$^{-2}$ at 1~keV per
$400\pi$~arcmin$^2$.
\end{tabnote}
\end{table*}

\begin{figure*}
\begin{center}
\includegraphics[width=80mm]{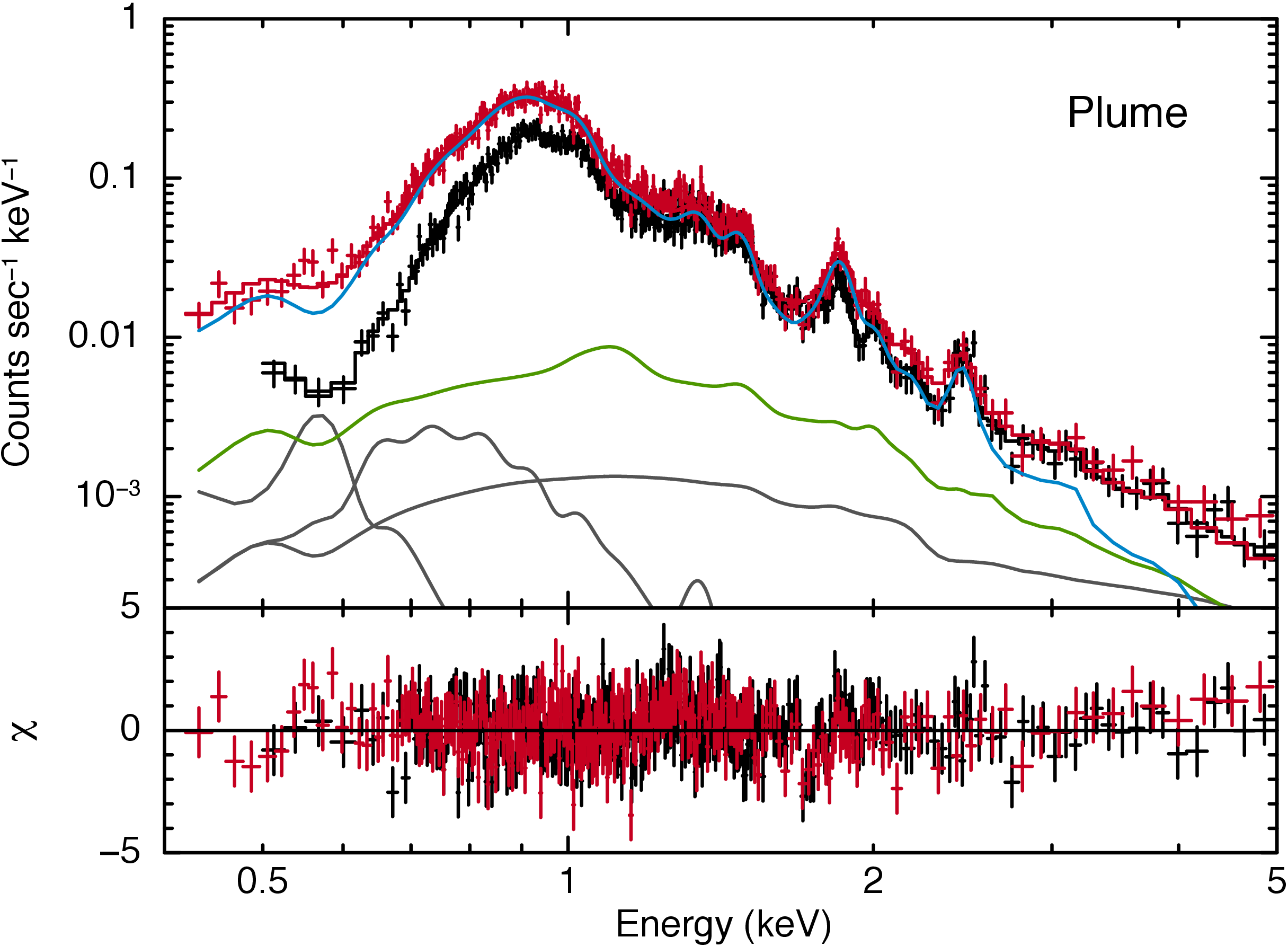}
\hspace{5mm}
\includegraphics[width=80mm]{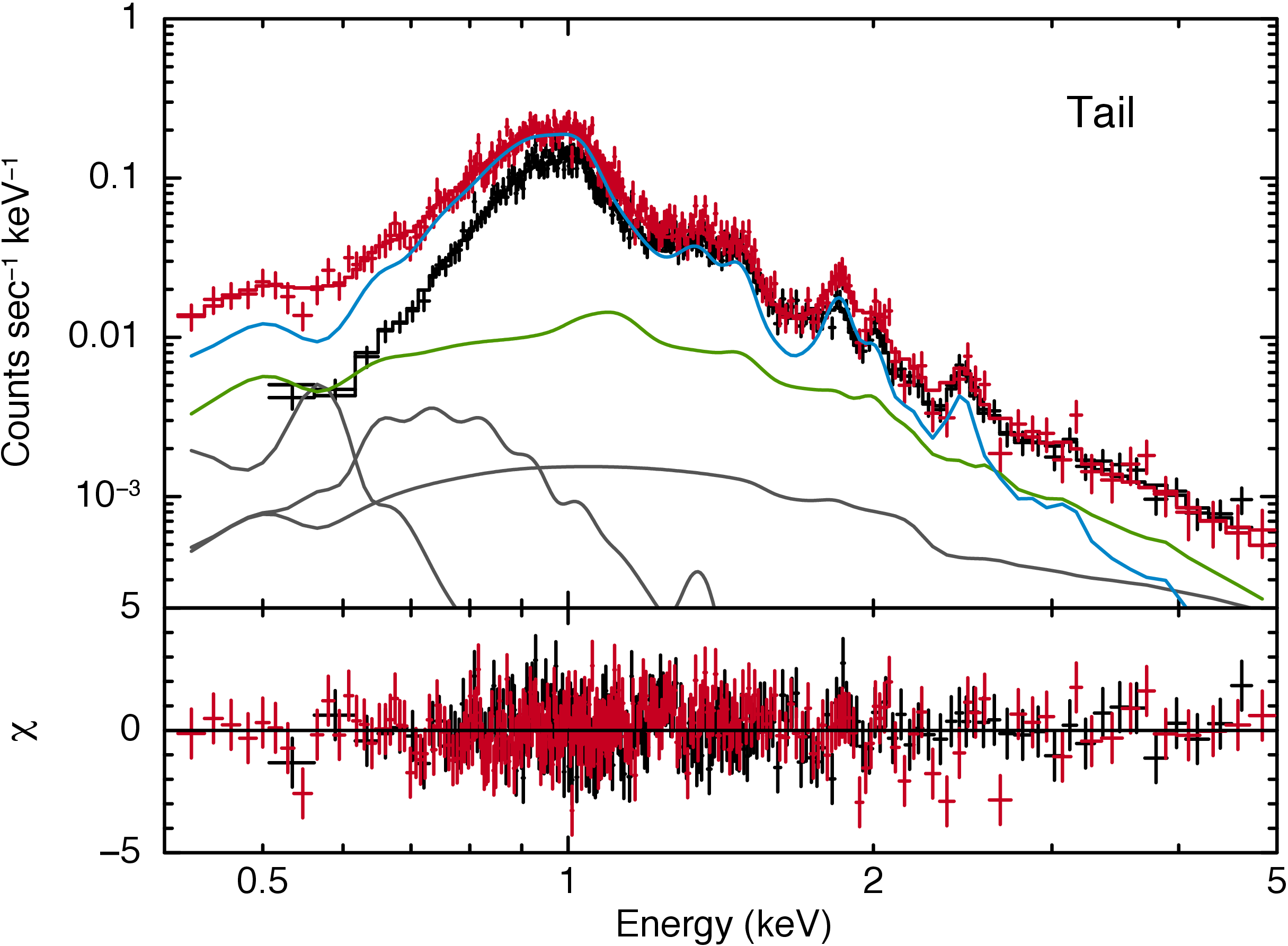}
\hspace{5mm}
\end{center}
\caption{Spectra of the Plume and Tail regions, and the best-fit
models. The red and black crosses show the data points of BI and FI CCD
data, and the red and black solid curves are the best fit models for
them. The blue, green, and gray curves are the vAPEC component, the ICM,
and the backgrounds/foreground components (CXB, LHB, MWH), respectively.
Only the components for the BI model are shown.}
\label{fig:plume_tail_spectra}
\end{figure*}

Thirdly, we fitted the spectra of the plume and the tail regions with
$1T$ model. The results are summarized in table~\ref{tab:plume_fit} as
case 1 (ICM temperature fixed at 2.1~keV) and case 2 (ICM temperature
free), and the best-fit models of case 1 are shown in
figure~\ref{fig:plume_tail_spectra}. The $F$-test probabilities between
case 1 and case 2 were $9.3\times10^{-3}$ and 0.20 for the plume and the
tail, respectively. The improvement was reasonable for the plume, but
the ICM temperature was $>3.2$~keV in case 2, which seemed too high as
the ICM temperature.  On the other hand, the abundances of the tail were
high ($\sim1.3Z_\odot$) when the ICM temperature was fixed at 2.1~keV
(case 1). The ICM temperature slightly higher than 2.1~keV is preferred
for both the plume and the tail.  When we employed a 2$T$ model, the
temperature of the second component became too high to
constrain. Therefore, $2T$ model was not meaningful for these regions.

A similar abundance pattern to that of the center region was seen in
both case 1 and 2, i.e., the abundances of O, Mg, Si, S, Fe were close
to each other, while that of Ne was about 2.5 times larger. 

\begin{figure*}
\begin{center}
\includegraphics[width=80mm]{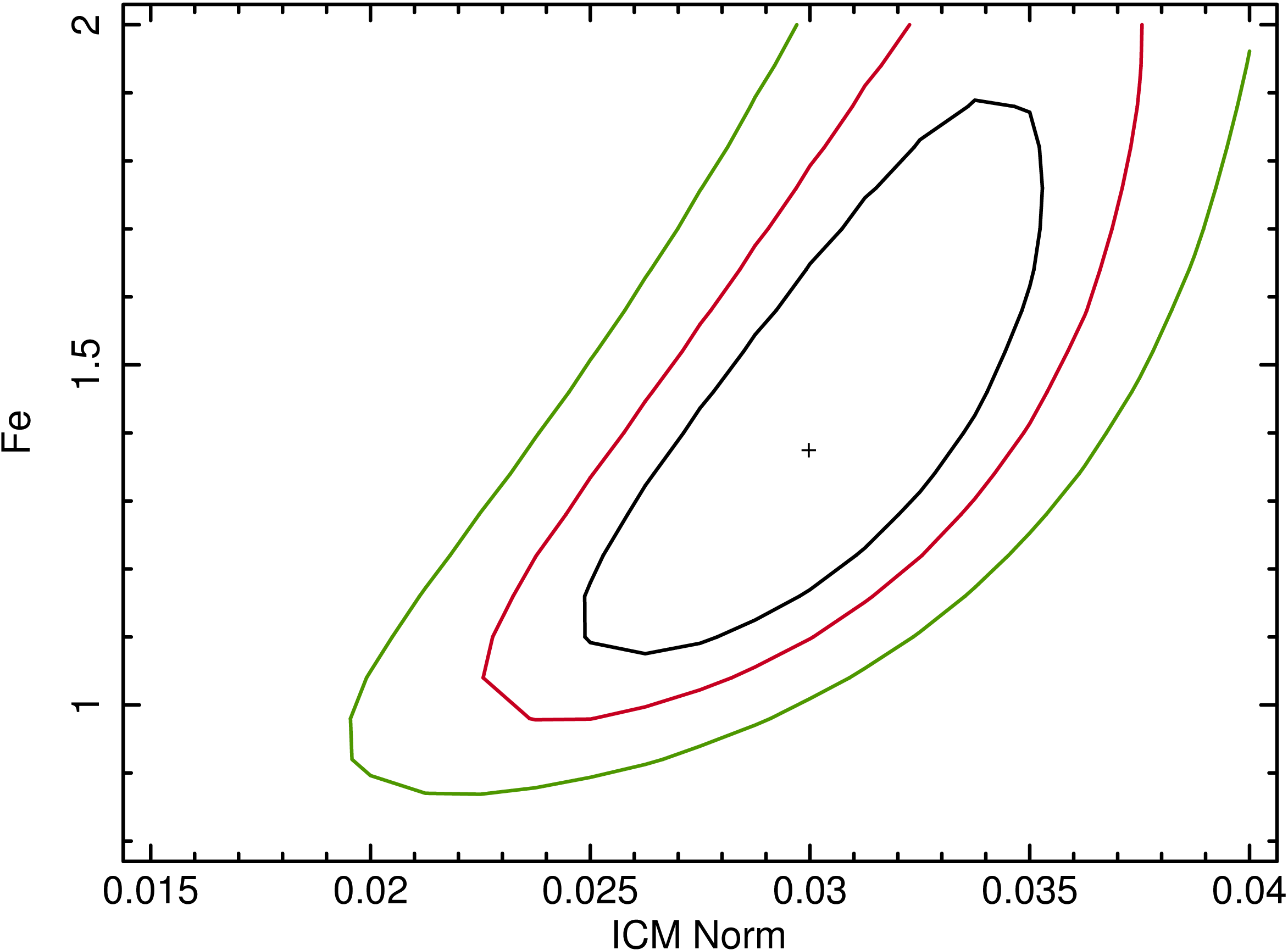}
\includegraphics[width=80mm]{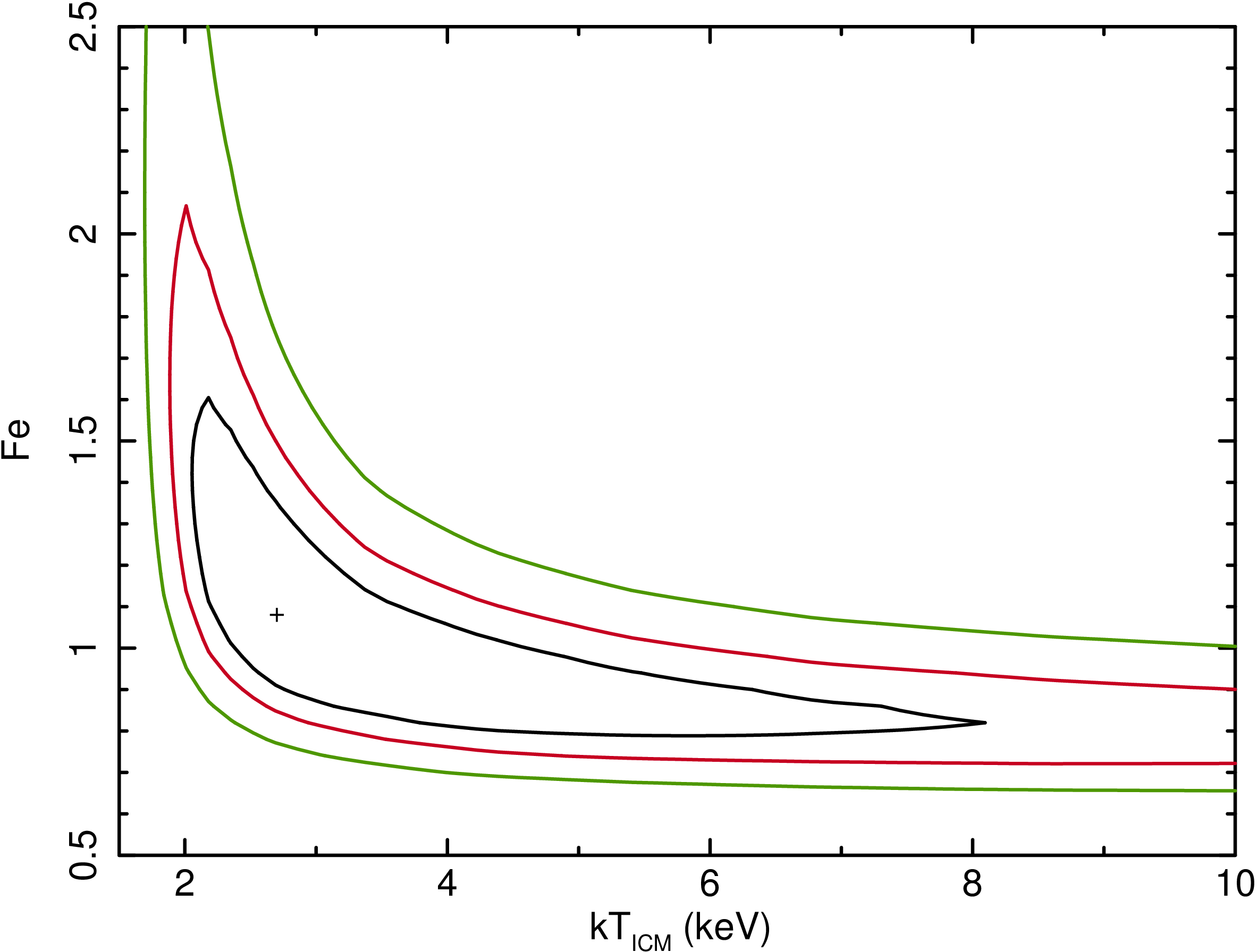}
\end{center}
\caption{(left) Confidence contour between Fe abundance and ICM
normalization, when the ICM temperature was fixed at 2.1~keV. (right)
That between Fe abundance and ICM temperature, when both the ICM
normalization and temperature were made free. The confidence levels are
$1\sigma$, 90\%, and $3\sigma$, respectively.}
\label{fig:contour_tail_Fe_ICMnorm}
\end{figure*}

The abundance of Fe, especially in the tail region, was slightly higher
than that of the center. In the tail region, the surface brightness was
relatively low, and the abundance (also the normalization) was affected
by the normalization and the temperature of the ICM.
Figure~\ref{fig:contour_tail_Fe_ICMnorm} shows a confidence contour
between the Fe abundance and the ICM normalization when the ICM
temperature was fixed at 2.1~keV, and a confidence contour between the
Fe abundance and the ICM temperature when both the ICM temperature and
normalization were free.  There is a positive correlation with the ICM
normalization and a negative correlation with the ICM temperature.  The
90\% lower limit of the Fe abundance is 0.9 if the ICM temperature is
2.44~keV. Therefore, we cannot conclude that the abundances in the tail
region is higher than those of the center region from the Suzaku data.

If we assume a uniform prolate spheroid of the equatorial radius of
\timeform{1.2'} and the polar radius of \timeform{2.7'} for the plume,
the hydrogen number density becomes $n_{\rm
plume}\approx6.4\times10^{-3}$~cm$^{-3}$ and the total mass is $M_{\rm
plume}\approx4.0\times10^{8}~M_\odot$. This is consistent with the
numbers obtained by \citet{M86_Chandra} within a factor of 2.  If we
assume a uniform cylinder of \timeform{1.7'} radius and \timeform{4.1'}
height for the tail region, the hydrogen number density becomes $n_{\rm
tail}\approx3.1\times10^{-3}$~cm$^{-3}$ and the total mass is $M_{\rm
tail}\approx4.5\times10^{8}~M_\odot$. The mass is about 1/4 of that
estimated by \citet{M86_Chandra}. Major difference is that our data only
covered part of the tail.

%\newpage
%%%%%%%%%%%%%%%%%%%%%%%%%%%%%%%%%%%%%%%%%%%%%%%%%%%%%%%%%%%%%%%%%%%%%%%%%
\section{Discussion}

\subsection{Core, plume, and tail}

As shown in the previous section, the temperatures of the core were
$kT=0.88^{+0.03}_{-0.04}$~keV and $\sim0.6$~keV ($2T$ fit), while those
of the plume and the tail were $0.86\pm0.01$~keV and $1.00\pm0.01$~keV,
respectively.  Thus, the temperature of the tail was slightly higher.
This is generally consistent with those reported by \citet{M86_Chandra}
and \citet{M86_XMM2}.  There was a tendency that the abundances became
slightly larger in the order of the core, the plume, and the
tail. However, we concluded that the difference was not significant,
thinking about statistical errors and also about systematic errors due
to variation of the ICM temperature and normalization.
\citet{M86_Chandra} pointed that the temperature structure of the tail
is consistent with a ram-pressure stripping model, i.e., the hotter,
higher entropy group gas is stripped first, followed by the cooler,
lower entropy M86 ISM. Our results strongly support it, since the
temperature of SE3 ($0.98^{+0.02}_{-0.03}$~keV) and the Fe abundance
(0.7--0.8$Z_\odot$) are close to those of the tail.

We determined the abundances of O, Ne, Mg, Si, S, and Fe separately, and
also showed that all the spectra of the different regions had a very
similar abundance pattern, i.e., O/Fe, Mg/Fe, Si/Fe, and S/Fe were
basically consistent with the solar ratio, while Ne/Fe was larger by a
factor of 3--4. \citet{Konami} analyzed Suzaku data of M86 as a whole,
i.e., including the core, the plume, and part of the tail together, and
reported that Ne/Fe was about 3. Our analysis showed that it was the
case with the center, the plume, the tail, and also the halo. The
spectra of these regions (figure~\ref{fig:fit_1T} and
\ref{fig:plume_tail_spectra}) showed a peak at around 0.9~keV and a hump
slightly above 1~keV.  They were caused by a forest of Fe L lines, and
Ne\emissiontype{X} Ly$\alpha$ lines at 1022~eV (Ne\emissiontype{X}
$2p\to 1s$), respectively, when the temperature was 0.8~keV.  The peak
energy due to the forest of Fe L lines rises as the temperature rises.
Therefore, the spectral shape in this energy region is mainly determined
by the temperature and the abundances of Fe and Ne.  The Ne abundance by
\citet{Lodders03} is about 60\% of that provided by \citet{AG89} or
\citet{GS98}. In the recent solar abundances provided by
\citet{Lodders10}, the Ne abundance is much closer to \citet{AG89} or
\citet{GS98}. Even if the new value were adopted, Ne overabundance would
be still significant, by a factor of 2--3. This common abundance
pattern, which is derived from the very similar spectral shapes, is one
evidence that the hot gas in the plume and the tail regions has the same
origin as that in the core.

% \begin{figure}
% \begin{center}
% \includegraphics[width=80mm]{figure13.ps}
% \end{center}
% \caption{Emissivities of Ne\emissiontype{X} Ly$\alpha$ lines at 1022~eV
% and Fe L lines in $1022\pm20$~eV energy range, as a function of a plasma
% temperature, based on AtomDB. Emissivity of Fe\emissiontype{XXI}
% $2p^13d^1\to 2p^2$ at 1009~eV is separately shown.}
% \label{fig:emissivity}
% \end{figure}

\citet{Konami} reported that the large Ne/Fe ratio cannot be explained
by any mixture of SNe type Ia and core-collapse SNe, and concluded that
Ne abundance may have intrinsically large systematic
errors because their emission
lines are hidden by prominent Fe-L lines. We further investigated it. 
To check model uncertainties, we fitted the center data with SPEX 3.0
\citep{SPEX}. When the temperature was fixed at the same
number, the abundances differed by $\sim30$\%. However, the difference
of Ne/Fe was only about 5\%, and hence, there was no significant
difference between the results based on APEC and SPEX.
According to AtomDB, there is an Fe L line (Fe\emissiontype{XVII}
$2p^54d^1\to2p^6$) at 1023 eV. In addition, there is a strong Fe L line
(Fe\emissiontype{XXI} $2p^13d^1\to 2p^2$) at 1009~eV, whose emissivity
is comparable to the summation of two Ne\emissiontype{X} Ly$\alpha$
lines when the temperature is $\sim0.8$~keV, and reaches the maximum 
when the temperature is $\sim1.1$~keV. 
Therefore, it may be difficult to determine the abundance of Ne
precisely in this temperature range unless Ne\emissiontype{X}
Ly$\alpha$ lines are separated from strong Fe L lines. Note that 
\citet{Ji09} showed high-resolution spectra of M86 obtained with
XMM-Newton reflection grating spectrometers (RGS), and the
Fe\emissiontype{XXI} $2p^13d^1\to2p^2$ line seemed to be
resolved. However, the neon abundance was not reported.

% Figure~\ref{fig:emissivity} shows the emissivities of
% Ne\emissiontype{X} Ly$\alpha$ lines, and Fe L lines in the
% $1022\pm20$~eV energy range. The energy range of $1022\pm20$~eV was
% selected since the energy resolution of the XIS is about 85~eV (FWHM) at
% around 1~keV. When the plasma temperature is 0.8~keV, emissivity of
% Fe\emissiontype{XXI} $2p^13d^1\to 2p^2$ at 1009~eV is comparable to that
% of Ne\emissiontype{X} Ly$\alpha$ lines, and the total emissivity of Fe L
% lines is about 3 times larger.
% % If the emissivities of Fe L lines are underestimated, the abundance of
% % Ne could become larger. 
% \textcolor{red}{Therefore, it may be difficult to determine the abundance of Ne
% precisely in this temperature region unless Ne\emissiontype{X}
% Ly$\alpha$ lines are separated from Fe L lines.}
% Note that the
% integrated Fe L line emissivity around 1022~eV starts increasing as the
% temperature rises, when the temperature is higher than $\sim0.7$~keV,
% while that of Ne\emissiontype{X} Ly$\alpha$ lines is decreasing.  This
% may explain why overabundance of Ne becomes significant when the ISM
% temperature is $\gtrsim0.7$~keV, as shown in figure~8 of \citet{Konami}.

%%%%%%%%%%%%%%%%%%%%%%%%%%%%%%%%%%%%%%%%%%%%%%%%%%%%%%%%%%%%%%%%%%%%%%%%%
\subsection{Extended halo}

\subsubsection{Characteristics of the halo gas}

The extended emission around M86 was clearly detected with
Suzaku. According to the XIS mosaic shown in figure~\ref{fig:image}, it
extends over \timeform{15'} (72~kpc) from the center in the east
direction, and over \timeform{10'} (48~kpc) in the south-west direction.
With the moderate spatial resolution of Suzaku, the surface brightness
profile of the core and the extended halo was represented with a single
$\beta$ model of $\beta\sim0.5$, as described in
section~\ref{sec:image_analysis}, and it indicated that the emission
spreads to $\sim\timeform{20'}$ ($\sim100$~kpc).  This picture is
consistent with the the spectral fit of SE6 region
($r=\timeform{16'}$--\timeform{20'}), which showed the existence of a
component of $kT=0.9$~keV.  If the surface brightness extends to a
certain distance following a $\beta$ model, the actual extent of the gas
must be significantly larger. Therefore, our results strongly suggest
that the halo of M86 extends over 100~kpc, at least in the east
direction.

Using the effective radius $r_e$ of \timeform{1.74'} \citep{RC3}, Suzaku
detected X-ray emission up to $\sim11.5r_e$ in SE1--6 regions. The ratio
of the temperature at 4--8$r_e$ to that at $<1r_e$ was
$kT(4{\rm-}8r_e)/kT(<1r_e)=1.21^{+0.05}_{-0.06}$, showing the positive
temperature gradient in the central region. \citet{Nagino09} denoted
galaxies with the temperature ratio $>1.3$ as X-ray extended galaxies
and others as X-ray compact galaxies. According to their criteria, M86
is located in the boundary area. The ratio of the stellar velocity
dispersion to the gas temperature $\beta_{\rm spec}\equiv\mu
m_p\sigma^2/kT$, where $\mu$ is the mean molecular weight in terms of
the proton mass $m_p$, is $\beta_{\rm spec}=0.47$, for the gas
temperature of 0.9~keV and the stellar velocity dispersion of
256~km\,s$^{-1}$ \citep{Roberts91}. This is close to the typical number
of the X-ray extended elliptical galaxies \citep{Matsushita01}.  These
results suggest that the halo gas is located in a larger scale potential
structure than that of the galaxy, such as a galaxy group
\citep{Matsushita01,Nagino09}.

\subsubsection{Estimation of the gas mass and the dynamical mass}

In section~\ref{sec:image_analysis}, we showed the surface brightness
distribution is represented by a $\beta$ model of $\beta\sim0.5$. In
this section, we estimate the gas mass assuming the $\beta$ model
obtained in section~\ref{sec:image_analysis}. The $\beta$ model was
valid from the position angle $\sim\timeform{45D}$ to
$\sim\timeform{275D}$, covering about 64\% of the whole area, and hence,
we used only this region.

In section~\ref{sec:core_halo_norm}, we estimated that the hydrogen
number density in the core was $(6$--$7)\times10^{-3}$~cm$^{-3}$,
assuming a uniform sphere of \timeform{1.5'} radius. We first calculated
the total emission measure along the line-of-sight through the M86
center ($r<\timeform{1.5'}$) assuming the $\beta$ model parameters shown
in table~\ref{tab:sb_fit}, and derived the density at $r=0$.  It became
(4.0--6.3)$\times10^{-3}$~cm$^{-3}$. In the following discussion, we
assume that the hydrogen number density at $r=0$ is
$5\times10^{-3}$~cm$^{-3}$.

If the $\beta$ model for the surface brightness distribution is valid to
infinity, the density is given by the following function:
\begin{equation}
n(r) = n_0 \left\{
1+\left(\frac{r}{r_0}\right)^2
\right\}^{-\frac{3}{2}\beta}.
\end{equation}
Since the hydrogen number density at the center is
$5\times10^{-3}$~cm$^{-3}$, the density at $r\sim100$~kpc is
$\sim2\times10^{-4}$~cm$^{-3}$. According to \citet{VC_XMM}, the
electron density of the Virgo ICM is $\sim2\times10^{-4}$~cm$^{-3}$ at about
350~kpc from the center of the Virgo cluster. Hence, the densities of
the halo gas and the ICM are comparable at around $r=100$~kpc.

\begin{figure}
\begin{center}
\includegraphics[width=80mm]{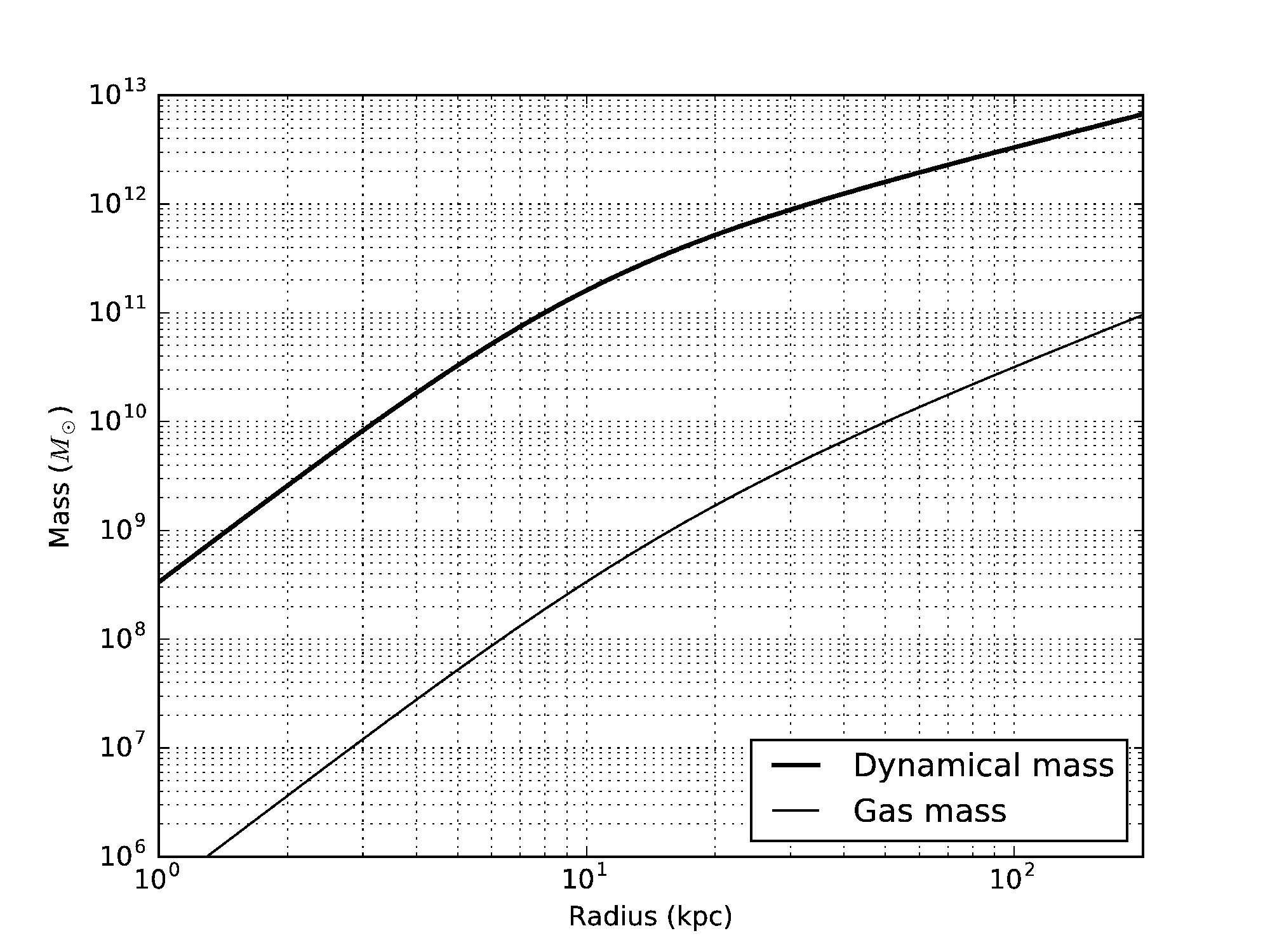}
\end{center}
\caption{Gas mass and dynamical mass from the position angle
\timeform{45D} to \timeform{275D}. Thin and thick curves correspond to
the gas mass and the dynamical mass, respectively.}  \label{fig:mass}
\end{figure}

The mass of the halo gas was estimated by
\begin{equation}
M_{\rm gas} = \sum_i \frac{\Delta\theta_i}{2\pi}\int_0^R 4\pi r^2\rho_i(r)dr,
\end{equation}
where $R$ is the radius of the gas, $\Delta\theta_i$ is the angle of the
$i$-th sector, $\rho_i(r)$ is the mass density at $r$ of the $i$-th
sector. The gas mass thus obtained is shown in figure~\ref{fig:mass} as
a function of radius. The mass of the halo gas in $r<100$~kpc became
$\sim3\times10^{10}M_\odot$.  

We further estimated the dynamical mass assuming the hydrostatic
equilibrium. The halo must be affected by the motion of M86 in the Virgo
cluster, and also by the ram-pressure stripping, and hence, the
hydrostatic equilibrium will not be a good approximation. However, we
think it is still useful to discuss the condition of M86. We calculated
the dynamical mass using the following equation:
\begin{equation}
 M_{\rm dyn}(r) 
= - \frac{kTr}{\mu m_H G}\left(\frac{d\ln\rho}{d\ln r} 
+ \frac{d\ln T}{d\ln r}\right)
\simeq \frac{3\beta kTr}{\mu m_H G}\frac{r^2}{r^2+r_0^2},
\end{equation}
assuming the temperature is almost constant. The dynamical mass thus
obtained is also shown in figure~\ref{fig:mass}.  It became
$\sim3\times10^{12}M_\odot$ for $r<100$~kpc. Then the ratio of the gas
mass to the dynamical mass $M_{\rm gas}/M_{\rm dyn}$ became $\sim0.01$.

\citet{Virgo_ROSAT} decomposed the X-ray surface brightness distribution
obtained with ROSAT into several components, and estimated the total
mass within 280~kpc is (1--3)$\times10^{13}M_\odot$. If we use their
number, $M_{\rm gas}/M_{\rm dyn}$ becomes only $\sim10^{-3}$.  If we
extend our calculation to 230~kpc (difference of the distance corrected)
assuming the same $\beta$ models are valid, the gas mass becomes
$\sim1\times10^{11}M_\odot$, and the ratio is
$\sim10^{-2}$. \citet{Schindler99} reported that the galaxy mass within
240~kpc from M86 center is $6\times10^{11}M_\odot$, and the ratio of the
galaxy mass to the total mass is 2--6\%. Therefore, depending on the
actual spread of the gas, the ratio of the gas mass to the galaxy mass
also significantly differs, from $\sim 0.1$ to $\sim 1$.

According to \citet{Sasaki15}, the gas mass fractions of clusters to the
hydrostatic mass are about 0.02--0.1. They also found that the ratio is
$10^{-3}$ for several subhalos in the Coma cluster, indicating
significant fraction of the gas was removed due to interaction with the
ICM, such as ram-pressure stripping. Our results implicate that the gas
mass to the dynamical mass ratio of M86 is $10^{-3}$--$10^{-2}$,
suggesting it is also significantly affected by the interaction with the
ICM.

\subsubsection{Timescales of Ram-pressure stripping in M86}

According to \citet{Forman79}, the ram-pressure stripping occurs when
the ram-pressure of the cluster gas exceeds the force holding the gas in
the galaxy:
\begin{equation}
 \rho_{\rm ICM}v^2 > \rho_{\rm ISM}\sigma_{\rm gal}^2
\end{equation}
where $\rho_{\rm ICM}$ and $\rho_{\rm ISM}$ are the ICM and ISM densities,
and $\sigma_{\rm gal}$ is the galaxy velocity dispersion (See also
\cite{Gunn_Gott}). At the core of M86, the ram-pressure is
$\sim6$~eV\,cm$^{-3}$ and $\rho_{\rm ISM}\sigma_{\rm gal}^2$ is
$\sim5$~eV\,cm$^{-3}$, for $n_{\rm ICM}=2\times10^{-4}$~cm$^{-3}$,
$n_{\rm ISM}=5\times10^{-3}$~cm$^{-3}$, $v=1500$~km\,s$^{-1}$, and
$\sigma_{\rm gal}=256$~km\,s$^{-1}$. Therefore, the ram-pressure
stripping condition is satisfied even in the core.

According to \citet{Frank92}, the mean free path $\lambda_\perp$ of the
Coulomb collisions of two protons that causes the large deflection
($\sim\timeform{90D}$) is given by
\begin{equation}
 \lambda_\perp \approx \frac{m_p^2v^4}{4\pi e^4n},
\end{equation}
and the time needed for the particle to travel the mean free path is
given by
\begin{equation}
 t_\perp = \frac{\lambda_\perp}{v}\approx\frac{m_p^2v^3}{4\pi e^4n}.
\end{equation}
They were calculated for the core and the halo and are shown in
table~\ref{tab:mfp}. The mean free path is comparable to the diameter of
the core and the halo, and hence, the close Coulomb collisions occur
with a large probability. In fact, there are much more distant
scatterings, and the mean free path and the travel time will be shorten
by a factor of the Coulomb logarithm ($\sim10$).

\begin{table}
  \tbl{Mean free path and travel time for close Coulomb collisions.}{%
  \begin{tabular}{lll}
\hline
 & Center & Halo \\
\hline
Number density $n_H$ (cm$^{-3}$)& $5\times10^{-3}$ & $2\times10^{-4}$\\
Mean free path $\lambda_\perp$ (kpc) & 8.6 & 215 \\
Travel time (y) & $5.6\times10^{6}$ & $1.4\times10^8$\\
\hline
\end{tabular}}\label{tab:mfp}
%\begin{tabnote}
%This is table note.
%\end{tabnote}
\end{table}

Time needed to strip all the halo gas is simply estimated by the total
number of the halo gas divided by the flux of the ICM, i.e.,
\begin{equation}
t_{\rm strip} \approx 
\frac{ \frac{4\pi}{3} n_{\rm ISM} R^3 }{4\pi n_{\rm ICM}R^2v}
=\frac{4}{3} \frac{n_{\rm ISM}}{n_{\rm ICM}}\frac{R}{v}.
\end{equation}
It becomes $\sim3\times10^8$~y in the core, and $\sim2\times10^8$~y in
the outer halo. Since the distance between M86 and M87 is 350~kpc in
projection, the crossing time is
\begin{equation}
 t_{\rm cross} \gtrsim \frac{2R}{v}\sim 4\times10^8\quad[{\rm y}].
\end{equation}
Therefore, $t_{\rm strip}\lesssim t_{\rm cross}$, and hence, most of the
gas in the core and in the halo will be stripped if M86 passes through
the Virgo cluster center once. 

We showed that the gas of low metal abundance still remains in the outer
halo.  On the other hand, it was indicated that the gas mass to the
dynamical mass ratio is $10^{-3}$--$10^{-2}$, which suggests significant
fraction of the halo gas has been stripped. The M86 group is experiencing 
the stripping by the Virgo ICM right now.

\subsection{X-ray clump near NGC~4388}
\label{sec:clump}

\begin{figure}
\begin{center}
\includegraphics[width=80mm]{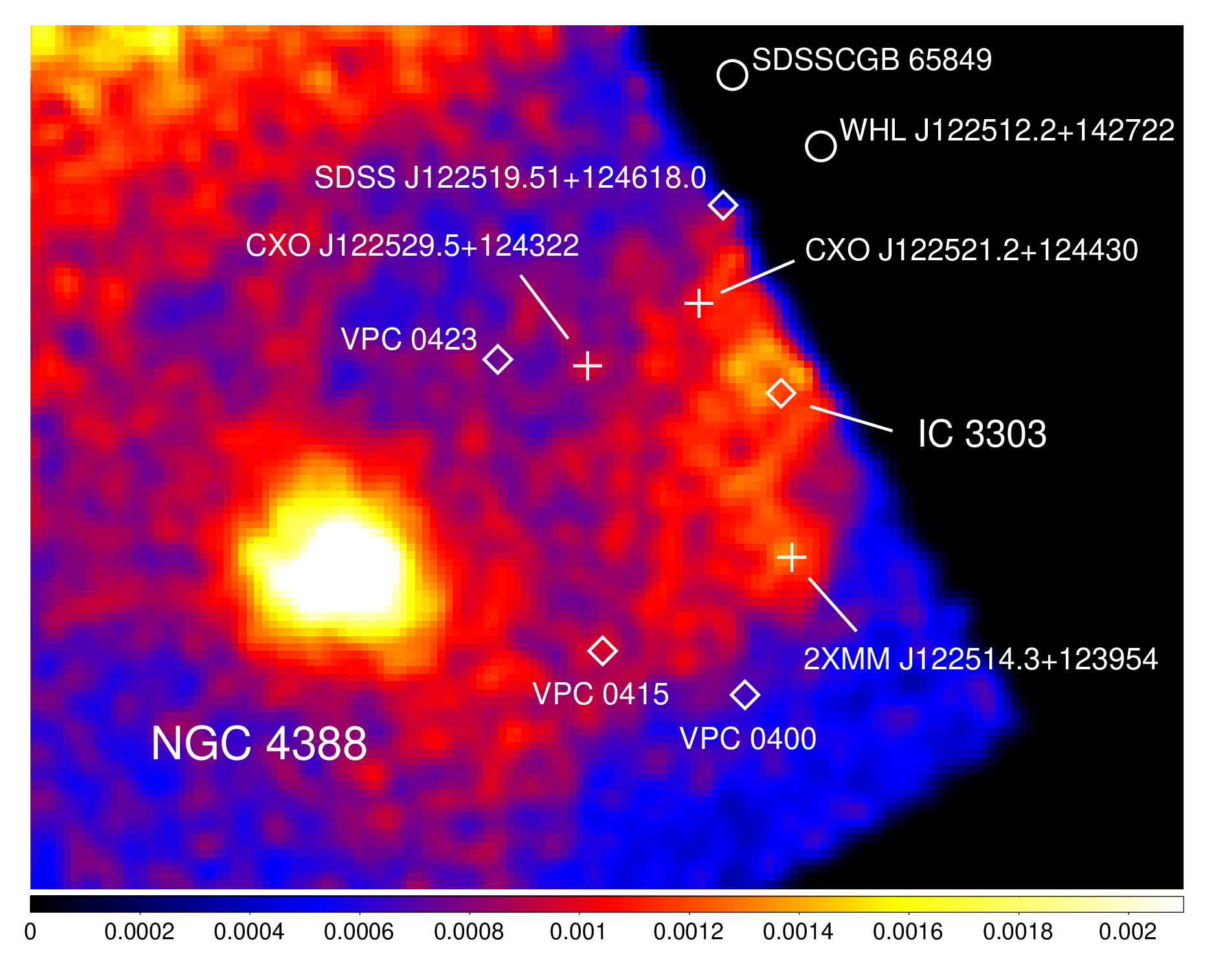}
\end{center}
\caption{XIS image around NGC 4388 and the X-ray clump. Circles, diamonds and
 crosses correspond to a cluster or a group of galaxies, galaxies with known
 redshift, and X-ray sources, respectively.}
 \label{fig:clump}
\end{figure}

As shown in section~\ref{sec:image_analysis}, a faint X-ray clump was detected near
NGC~4388. The temperature of the gas was $\sim1$~keV and its flux in the
0.5--2~keV band was $\sim 4 \times 10^{-12}$~erg\,cm$^{-2}$\,s$^{-1}$. We
could not find any
literature mentioning it. We checked the NASA/IPAC Extragalactic
Database (NED)\footnote{http://ned.ipac.caltech.edu/.} to see if there
are any associations with known background groups or clusters. One
cluster WHL~J122512.2$+$142722 at $z=0.4264$ and one group SDSSCGB~65849
were found near the XIS field of view, but it is unlikely that either of
them is an optical counterpart. Next, we looked for galaxies with known
redshift and X-ray sources in NED. The results are summarized in
figure~\ref{fig:clump}. Among them, IC~3303 is located near the
brightest part of the
X-ray clump. The redshift of this galaxy is $-0.000627\pm0.000077$
\citep{IC3303}, which is very close to that of
M86. Therefore, this clump might be part of the M86 subgroup gas, though
it is separated from the extended emission around M86. Further study is
needed.

%%%%%%%%%%%%%%%%%%%%%%%%%%%%%%%%%%%%%%%%%%%%%%%%%%%%%%%%%%%%%%%%%%%%%%%%%
\section{Summary}

We analyzed the Suzaku data of M86 and its adjacent regions to study the
extended emission around it. The M86 core, the plume, and the tail
extending toward the northwest were clearly detected, as well as the
extended halo around them. From the position angle $\sim\timeform{45D}$
to $\sim\timeform{275D}$, the surface brightness distribution was
represented relatively well with a single $\beta$-model of
$\beta\sim0.5$ up to \timeform{15'}--\timeform{20'}.

The Suzaku XIS spectra of the M86 center, the extended halo, the plume,
and the tail were explained with one- or two-temperature plasma model,
in addition to the Virgo ICM of $kT\approx2.1$~keV and other
background/foreground components. The temperatures of the center were
$0.88^{+0.03}_{-0.04}$~keV and $\sim0.6$~keV. The temperatures of the
core and the halo have a positive gradient, and reach the maximum of
$kT\sim1.0$~keV at $r\sim\timeform{7'}$ or $\sim4r_e$. Outside it, it is
almost constant or slightly decreasing toward the outer regions. The
temperature of the plume and the tail were $0.86\pm0.01$~keV and
$1.00\pm0.01$~keV, respectively. Therefore, the temperature of the tail
is slightly higher than the core and the plume. These were qualitatively
consistent with the previous Chandra and XMM-Newton results
\citep{M86_Chandra,M86_XMM2}.

We succeeded in determining the abundances of O, Ne, Mg, Si, S, and Fe
separately, for the core, the plume, the tail, and the halo for the
first time.  The best-fit values of the Fe abundance in the core and
in the plume were $\sim0.7$, while that in the tail was slightly higher
($\sim1.0$). However, we cannot conclude that the abundance in the tail
is higher, thinking about the normalization and the temperature
variation of the ICM. The abundance of the halo is almost the same up to
$\sim\timeform{10'}$, and then it becomes significantly smaller
(0.2--0.3) at $r\gtrsim\timeform{10'}$. This means that the gas in the
outer halo is less polluted by the metals produced in the galaxy. In all
the regions, the abundance ratios of O, Mg, Si, and S to Fe were
$\sim1$, while Ne/Fe showed a significantly larger number (2--4). This
Ne overabundance is coming from the spectral features at around 1~keV,
and is another evidence that the plume and the tail have the same origin
as the core. However, the overabundance by a factor of 2--4 cannot be
explained by the uncertainty of the abundances, or mixture of known SNe
nucleosynthesis models.
Ne abundance may have intrinsically large systematic errors
as suggested by \citet{Konami}.

Our results suggest that the halo of M86 extends over 100~kpc, at least
in the east direction. The temperature at the center is slightly lower,
and the ratio of the stellar velocity dispersion to the gas temperature
is only 0.47. These features indicate that the extended halo gas is
located in a larger scale potential structure than that of the galaxy,
such as a galaxy group \citep{Nagino09,Matsushita01}. Using the $\beta$
models for sectors, we estimated the gas mass from the position angle
$\sim\timeform{45D}$ to $\sim\timeform{275D}$ (64\% of the whole
area). It was $\sim3\times10^{10}M_\odot$ in $r<100$~kpc. If we further
assume the hydrostatic equilibrium, the dynamical mass in the same
region was $\sim3\times10^{12}M_\odot$, giving the ratio of the gas mass
to the dynamical mass $M_{\rm gas}/M_{\rm dyn}\sim0.01$. If we adopt the
dynamical mass within 230~kpc provided by \citet{Virgo_ROSAT}, the ratio
becomes $\sim10^{-3}$. These ratios suggest the halo of M86 is
significantly affected by the interaction with the Virgo ICM.  Simple
estimation of the ram-pressure stripping lengthscales and timescales
showed that the mean free path is comparable to the size of the core or
the halo, and the stripping timescale is comparable or shorter than the
crossing time through the Virgo center. Therefore, most of the gas in
the core and in the halo will be stripped if M86 passes through the
Virgo center once. The fact that the low metal gas still remains in the
outer halo indicates that the M86 group is experiencing the stripping by
the Virgo ICM right now.

%\clearpage
%%%%%%%%%%%%%%%%%%%%%%%%%%%%%%%%%%%%%%%%%%%%%%%%%%%%%%%%%%%%%%%%%%%%%%%%%
\begin{ack}
We thank Prof. Kyoko Matsushita for her valuable comments on X-ray
 properties of the elliptical galaxies and interaction with the ICM. We
 also acknowledge Dr. Toru Sasaki, who supported our analysis. We are
 grateful to the anonymous referee, who provided useful comments to
 improve our manuscript.
\end{ack}

%\appendix 
%\section*{Effect of finite extent of density to the surface brightness}
%\label{sec:finite_extent}
%
%\section{Case of two or paragraphs}
%
%\section{Case of two or paragraphs}

%%%
% See the manual for the detail.
%%%

\end{document}